%% file: distinctive-points.tex
\pdfoutput=1
\documentclass[acmtog]{acmart}

\usepackage{booktabs} % For formal tables
 
\usepackage{multirow}

\usepackage[ruled]{algorithm2e} % For algorithms

\SetAlFnt{\small}
\SetAlCapFnt{\small}
\SetAlCapNameFnt{\small}
\SetAlCapHSkip{0pt}

\acmJournal{TOG}
\begin{document}
	
\newcommand{\TODO}[1]{{\color{red}{[TODO: #1]}}}
\newcommand{\dc}[1]{{\color{blue}#1}}
\newcommand{\phil}[1]{{\color[rgb]{0,0,0}#1}}
\newcommand{\xz}[1]{{\color[rgb]{0,0,0}#1}}
\newcommand{\lqyu}[1]{{\color{cyan}#1}}
\newcommand{\NEW}[1]{{\color{black}#1}}
\newcommand{\NEWNEW}[1]{{\color{black}#1}}

\newcommand{\red}[1]{{\color[rgb]{1,0,0}#1}}

\newcommand{\pos}{\mathbf{p}}
\newcommand{\posmat}{\mathbf{P}}
\newcommand{\para}[1]{\vspace{.05in}\noindent\textbf{#1}}
\def\ie{\emph{i.e.}}
\def\eg{\emph{e.g.}}
\def\etal{{\em et al.}}
\def\etc{{\em etc.}}
	
% Title portion
\title{Unsupervised Detection of Distinctive Regions on 3D Shapes}

\input{abstract}

% DO NOT ENTER AUTHOR INFORMATION FOR ANONYMOUS TECHNICAL PAPER SUBMISSIONS TO SIGGRAPH 2019!
\author{Xianzhi Li}
\author{Lequan Yu}
\author{Chi-Wing Fu}
%\orcid{1234-5678-9012-3456}
\affiliation{%
  \institution{The Chinese University of Hong Kong}
  %\streetaddress{104 Jamestown Rd}
  %\city{Williamsburg}
  %\state{VA}
  %\postcode{23185}
  %\country{USA}
}
\email{{xzli,lqyu,cwfu}@cse.cuhk.edu.hk}
  
\author{Daniel Cohen-Or}
%\orcid{1234-5678-9012-3456}
\affiliation{
 	\institution{Tel Aviv University}
 	%\streetaddress{104 Jamestown Rd}
 	%	\city{Tel Aviv}
 	%\state{China}
 	%\postcode{23185}
 	%	\country{Israel}
}
\email{dcor@mail.tau.ac.il}
  
\author{Pheng-Ann Heng}
%\orcid{1234-5678-9012-3456}
\affiliation{
 	\institution{The Chinese University of Hong Kong}
 	%\streetaddress{104 Jamestown Rd}
 	%	\city{Hong Kong}
 	%\state{China}
 	%\postcode{23185}
 	%	\country{China}
}
\email{pheng@cse.cuhk.edu.hk}

\renewcommand\shortauthors{Li X. et al.}




\maketitle

\input{introduction}
%\input{introduction2}
\input{related_work}

%\input{problem_statement}
\input{method2}

%\input{visualization}

\input{experiment}
\input{applications}

\input{conclusion}

\begin{acks}
	We thank all the anonymous reviewers for their comments and
	feedback. We also acknowledge help from our volunteers for conducting user studies. This work was supported by grants from the Research Grants Council of the Hong Kong Special Administrative Region (Project no. CUHK 14201717 and 14201918), the CUHK Research Committee Direct Grant for Research 2018/19, and the Israel Science Foundation as part of the ISF-NSFC joint program (grant number 2217/15, 2472/17). The work was also partially supported by ISF grant 2366/16.
\end{acks}

\bibliographystyle{ACM-Reference-Format}
%\nocite{*}
\bibliography{distinctive-points}

\end{document}

%% file: abstract.tex
\begin{abstract}
This paper presents a novel approach to learn and detect distinctive regions on 3D shapes.
Unlike previous works, which require labeled data, our method is unsupervised.
We conduct the analysis on point sets sampled from 3D shapes, then formulate and train a deep neural network for an unsupervised shape clustering task to learn local and global features for distinguishing shapes with respect to a given shape set.
To drive the network to learn in an unsupervised manner, we design a clustering-based nonparametric softmax classifier with an iterative re-clustering of shapes, and an adapted contrastive loss for enhancing the feature embedding quality and stabilizing the learning process.
By then, we encourage the network to learn the point distinctiveness on the input shapes.
We extensively evaluate various aspects of our approach and present its applications for distinctiveness-guided shape retrieval, sampling, and view selection in 3D scenes.
%We extensively evaluate various aspects of our approach and present its applications for distinctiveness-guided shape retrieval and sampling, as well as distinctive region detection in 3D scenes.
%%
\end{abstract}

%% file: introduction.tex
\section{Introduction}
\label{sec:introduction}

Reasoning about distinctive regions on 3D shapes has a wide range of applications in computer graphics and geometric processing,~\eg, object retrieval~\cite{shilane2006selecting,gal2006salient}, shape matching~\cite{castellani2008sparse,shilane2007distinctive}, and view selection~\cite{lee2005mesh,leifman2012surface}.
%shape simplification~\cite{howlett2005predicting}, 
%
In this work, we follow the definition of \emph{distinctive} regions proposed by Shilane and Funkhouser~\shortcite{shilane2006selecting,shilane2007distinctive},~\ie, \NEW{the distinction of a surface region in an object is defined as}
\begin{quote}
\NEW{\em 
how useful the region is for distinguishing the object from others of different types\footnote{\NEWNEW{Type refers to the specific class that an object belongs to,~\eg, chair, table, and car.}}.}
\end{quote}
\NEW{Hence, distinctive regions of a shape should be \emph{common and unique} in its own type, compared with shapes of other types.}
%regard a region on a 3D shape as {\em distinctive\/} if the region {\em helps distinguish the shape from shapes of other types\/}.
%
So, distinctive involves, and is quantified {\em relative to\/}, a given set of 3D shapes.

\begin{figure}[t]
	\centering
	%\vspace*{-3mm}
	\includegraphics[width=0.99\linewidth]{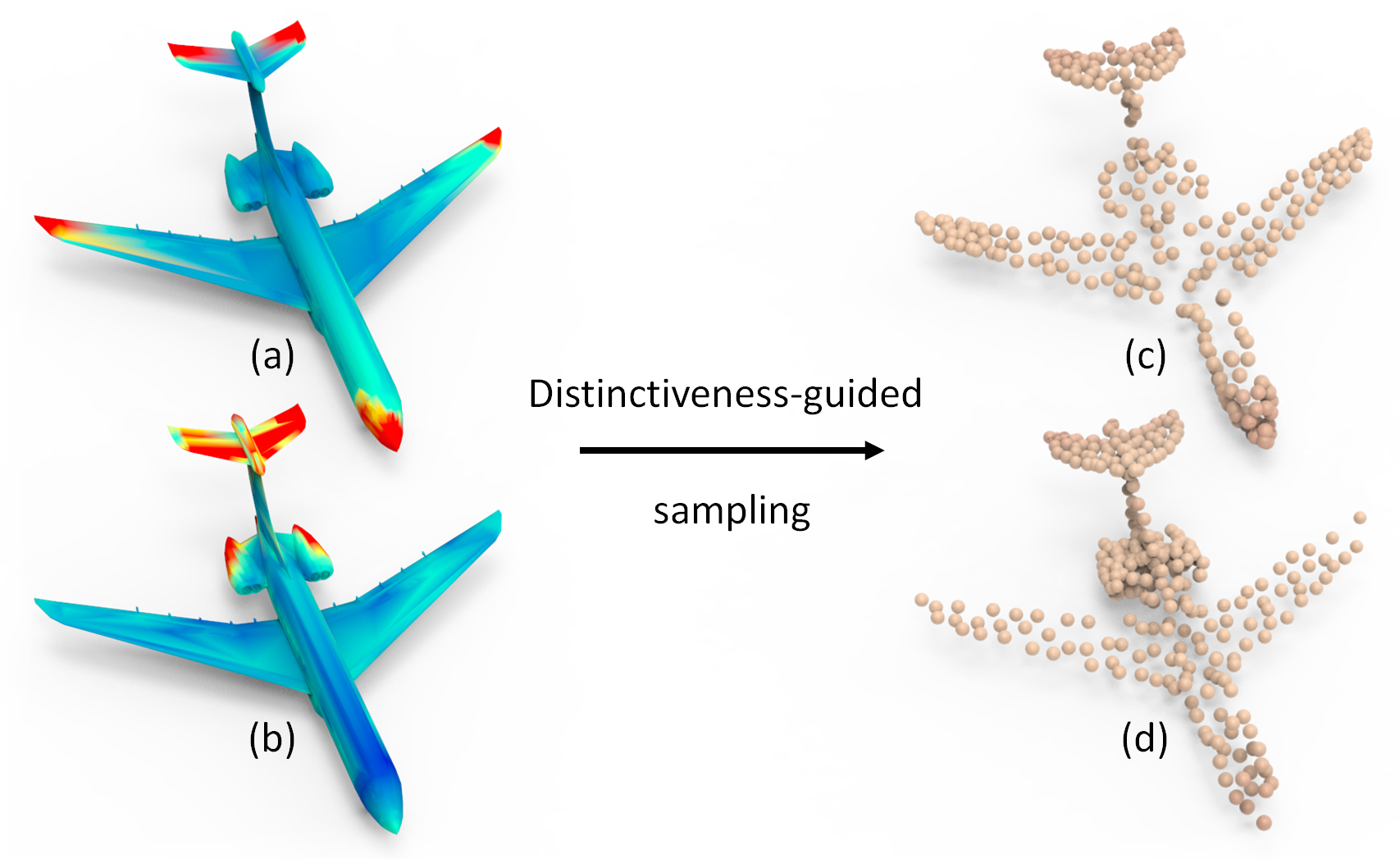}
	\vspace*{-2mm}%\phil{compensate the white space}
	\caption{Distinctive regions detected by our method \NEW{on the same shape shown in (a,b), relative to}
	(a) an inter-class dataset \NEW{(i.e., the whole ModelNet40 dataset)} and
	(b) an intra-class dataset \NEW{(i.e., only the tail-engine and four-engine airplanes in ModelNet40)},
	with their corresponding distinctiveness-guided sampling results (c,d).
	The colors in (a,b) indicate the region distinctiveness with red color being the most distinctive.}
	%\NEW{The inter-class dataset is the whole ModelNet40, while the intra-class dataset comprises only tail-engine and four-engine airplanes in ModelNet40.}}
	\label{fig:teaser1}
	\vspace{-2mm}
\end{figure}
\begin{figure}[t]
	\centering
	%\vspace*{-3mm}
	\includegraphics[width=0.99\linewidth]{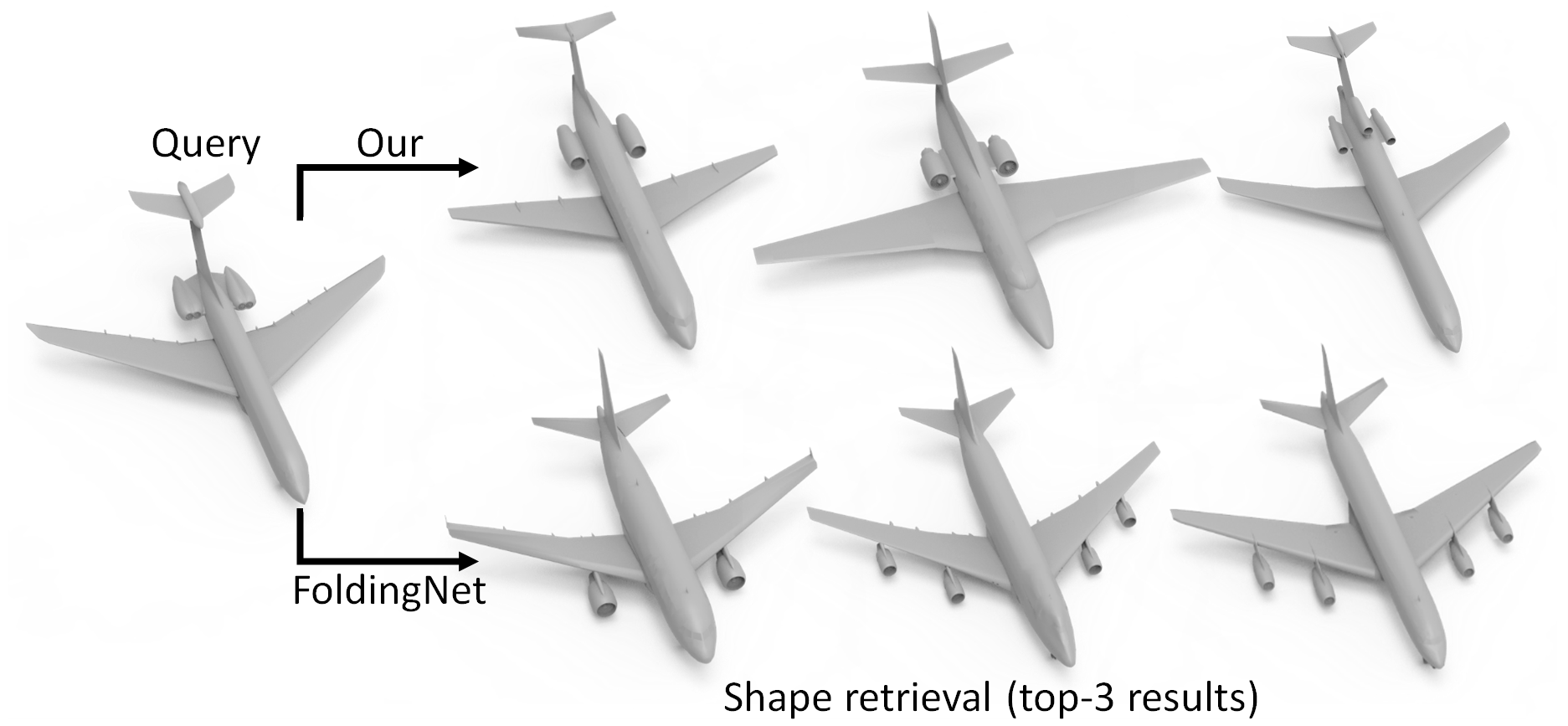}
	\vspace*{-2mm}%\phil{compensate the white space}
	\caption{Top-3 Shape retrieval results using our intra-class distinctiveness-guided retrieval method (top row) and using FoldingNet~\cite{yang2018foldingnet}, a general unsupervised feature learning method (bottom row).
	Conventional methods can only retrieve shapes with similar overall appearance, while {\em guided by distinctiveness\/} enables the retrieval of shapes with similar distinctive features,~\eg, the two engines at the back.}
	\label{fig:teaser2}
	\vspace{-3.5mm}
\end{figure}

However, existing methods~\cite{shilane2006selecting,shilane2007distinctive,song2018distinction} either rely on hand-crafted local features, or detect distinctive regions in a supervised setting, meaning that they require labels on data.
%besides hand-crafted local features, existing methods~\cite{shilane2006selecting,shilane2007distinctive} detect distinctive regions in a supervised setting, meaning that they require labels on data.
% and often assume that the given shape set has been pre-classified.\lqyu{what is the meaning of pre-classified? Does it have a same meaning with `require labels'?}
%
%First, perceptual measures are usually influenced by both local and global features. CNN-based methods have demonstrated that they are good at extracting such features and making good balance between them in the tasks of computing perceptual measures for images [10], [11].
%%
Hand-crafted features generally do not generalize well to other shapes, since their representation capabilities are limited by the pre-defined fixed operations.
%%
%In contrast, CNNs learn features specific to the data and have demonstrated strong generalisation capability through various applications.
%%
\NEW{Also,} in most application scenarios, it is difficult to acquire labels or pre-classify 3D shapes due to annotation efforts.
In light of these limitations, the challenging problem is to explore unsupervised methods to detect distinctive regions directly from the 3D shapes in a data-driven manner.
In such a setting, the set of given shapes has {\em not\/} been pre-analyzed by any means, 
%not even weakly supervised, 
\NEW{and no local descriptors are pre-defined on them.}

In this work, we present a method to compute distinctive regions on 3D shapes {\em in an unsupervised setting\/}.
Our method is based on a neural network that learns and analyzes a given set of shapes without relying on hand-crafted local features, and assigns to each point on the shapes a degree of distinctiveness.
%Based on these distinctiveness degree, we can further locate distinctive points and regions.
%
First, we sample and represent each given shape as a point cloud, a lightweight and flexible representation.
We design a deep neural network and train it on the point clouds for an \textit{unsupervised clustering task}.
By training, our network can learn both per-point features and per-shape features.
% assign a local feature to each point in the shapes and a global one to each shape.
%
In particular, to drive the network to learn to cluster the shapes for detecting distinctive regions, we design a joint loss function composed of a clustering-based nonparametric softmax loss and an adapted contrastive loss.
%To enhance the learning in our unsupervised setting, we design a joint loss function composed of a clustering-based nonparametric softmax loss and an adapted contrastive loss.
%%
For the network to learn to cluster the shapes, it has to attend to the discriminative features among the shape clusters.
Hence, by analyzing the resulting per-point features and per-shape features, we can obtain a degree of distinctiveness per point in the point sets, and further project the per-point distinctiveness in each point set back to the original shape.

Figure~\ref{fig:teaser1} shows two visual examples of using our method to find distinctive regions \NEW{in the same shape} (red being the most distinctive) relative to two different datasets:
%an intra-class dataset (a) and an inter-class dataset (b), respectively.
\NEW{the inter-class dataset (a) is the whole ModelNet40 dataset~\cite{wu20153d}, whereas the intra-class dataset (b) comprises only the tail-engine and four-engine airplanes in ModelNet40.}
%different sets of shapes.
%
%For both results, the network succeeds to recognize, or classify, the shape mainly by considering the distinctive regions.\lqyu{??}
%, \phil{since we train the network for a shape clustering task.}
%These points are said to be the most distinctive in the ``eyes'' of the network.
%\NEW{Note that, the intra-class dataset contains only tail-engine airplanes and four-engine airplanes, while the inter-class dataset contains not only non-airplane shapes, but also various kinds of airplanes.}
%
Comparing \NEW{the distinctive regions in (a) and (b)} reveals an interesting phenomenon that (a) tends to focus on the \NEW{contour of the airplane}, while (b) tends to pay more attention to the local regions.
\NEW{This result corresponds to the definition of distinction, since in the inter-class dataset, the contour of the tail-engine airplane is common and unique in the airplane class compared with others, while in the intra-class dataset, the tail and engines are more helpful for distinguishing the tail-engine airplanes from the four-engine airplanes.}
%This result corresponds to our intuition that to distinguish whether a given shape is an airplane needs to focus on the overall shape, while to distinguish among different types of airplane requires to focus more on the details,~\eg, the engines.
%
%By sampling more points on the distinctive regions (c) \& (d), even the point number is very small, there has little effect on the shape classification.
Sampling the point sets away from the distinctive regions only has little effect on the classification (see Figure~\ref{fig:teaser1} (c) \& (d)).
Later, we shall show, in an extensive empirical experiment presented in Section~\ref{subsec:distinctive}, that the points located on the distinctive regions are the {\em key points\/} for classification performance.
Besides, the detected distinctive regions can further facilitate the development of various applications, e.g., 
%high-precision
fine-grained shape retrieval; see an example result in Figure~\ref{fig:teaser2}, and Sections~\ref{sec:experiment} and~\ref{sec:applications} for more results.
%The retrieved results are all the same type (~\ie, tail-engine) as the query; see the middle row in Figure~\ref{fig:teaser}.
%%
%However, a general unsupervised global feature learning method,~\ie, FoldingNet~\cite{yang2018foldingnet}, has no ability to retrieve \emph{locally}similar objects; see the bottom row.

\NEW{It is worth noting that, the notion of distinction is closely related to saliency as they both measure regional importance. However, while distinction considers how common and unique a region is relative to objects of other types, saliency considers how unique and visible a region is relative to other regions within the same object.}

Overall, the contributions of this work are summarized below.
%
%First,
We develop a novel unsupervised framework to detect distinctive regions on 3D shapes that does not require hand-crafted features and labels on data.
%
%Second, we introduce a novel clustering-based softmax classifier to guide the network for unsupervised training.
%Second, we also present a network with a novel clustering-based softmax classifier for unsupervised deep feature learning.
%Second, w
We design a new clustering-based nonparametric softmax classifier and adopt an adapted contrastive loss to encourage the network to learn in an unsupervised manner.
%
%Lastly, we demonstrate the utility of our shape distinction for adaptive sub-sampling and object retrieval. 
%Lastly, 
We performed \NEW{extensive} experiments to evaluate the effectiveness of our method: 
quantitatively evaluating on how the detected distinctive regions help shape classification, 
%comparison with other methods,
a user study to compare our results with human, etc.
Further, we show how distinctiveness contributes to applications for shape retrieval, sampling, and view selection.
%in 3D scenes.
%object retrieval, distinctiveness-driven re-sampling, and distinctive region detection in 3D scenes.

%\phil{******NOTE****** I tried to fit Sec 1 & 2 on the first two pages as well}

%% file: related_work.tex
\section{Related work}
\label{sec:rw}
%In this section, we first briefly review techniques for detecting important regions (or points) of 3D objects. 
%We then give a partial overview of unsupervised learning in natural images.
%Finally, we present some works for point sets deep learning.

%%%%%%%%%%%%%%%%%%%%%%%%%%%%%%%%%%%%%%%%%%%%%%%%%%%%%%%%%%%%

%\vspace*{-3mm}
\paragraph{Distinctive region detection.}
%\subsection{Distinctive region detection}
%
The concept of \emph{distinction}, or \emph{distinctiveness}, was first proposed by Shilane and Funkhouser~\shortcite{shilane2006selecting,shilane2007distinctive}.
%In their method, shape descriptors are first computed for various regions on each shape.
%Then, the descriptors are mapped into a space parameterized by their likelihood for evaluating the distinctiveness of each region based on the retrieval performance on a training set of labeled descriptors.
The main idea of their methods is to extract local shape descriptors for local regions on each shape, then to obtain the distinctiveness of each local region by comparing the difference between all pairs of shape descriptors in the training database.
To avoid the drawback of hand-crafted shape descriptors, Song~\etal~\shortcite{song2018distinction} employed a classification network to consume multi-view images of given 3D shapes as input and learn view-based distinction by back-propagating the classification probability.
Next, a Markov random field is employed to combine the view-based distinctions across multiple views.
%\xz{
%As a view-based approach, it is not entirely invariant to object orientation.
%%
%%To avoid the drawback of hand-crafted shape descriptors, Song~\etal~\shortcite{song2018distinction} developed a method based on a classification network and a Markov Random Field (MRF). 
%%The classification network consumes multi-view images of given 3D shapes as input and learns view-based distinction by back-propagating the classification probability; then, the MRF is used to combine the view-based distinctions across multiple views.
%%
%%Despite the promising results they produced,
Despite the success in finding distinctive regions, existing approaches are {\em all\/} supervised, meaning that they all need class labels on the shapes given in the training dataset.
In contrast, our method detects distinctive regions in an unsupervised manner. 

Besides 3D shapes, the concept of \emph{distinction} was also mentioned in several works on images.
%\cite{lowe2004distinctive} presented a method to extract distinctive features that can be applied for image matching between different views.
%
Given a large collection of geo-localized images,
Doersch~\etal~\shortcite{doersch2012makes} developed a discriminative clustering approach to find visual elements that occur much more often in one geographic region than in others,~\eg, the kinds of windows, balconies, and street signs that are distinctive in Paris, compared with those in London.
%the most distinctive for a certain geo-spatial area \phil{such as Paris? what else?}
%\phil{can you say more? did you compare image set to image set?}.
%\phil{what is their definition of discriminative?}
%
Later, several approaches were developed to extract discriminative regions from images for image classification~\cite{singh2012unsupervised,juneja2013blocks,sun2013learning}.
More recently, Wang~\etal~\shortcite{wang2016mining} proposed a patch-based framework by introducing triplets of patches with geometric constraints to mine discriminative regions for fine-grained intra-class classification.

\begin{figure*}[t]
	\centering
	\includegraphics[width=0.95\linewidth]{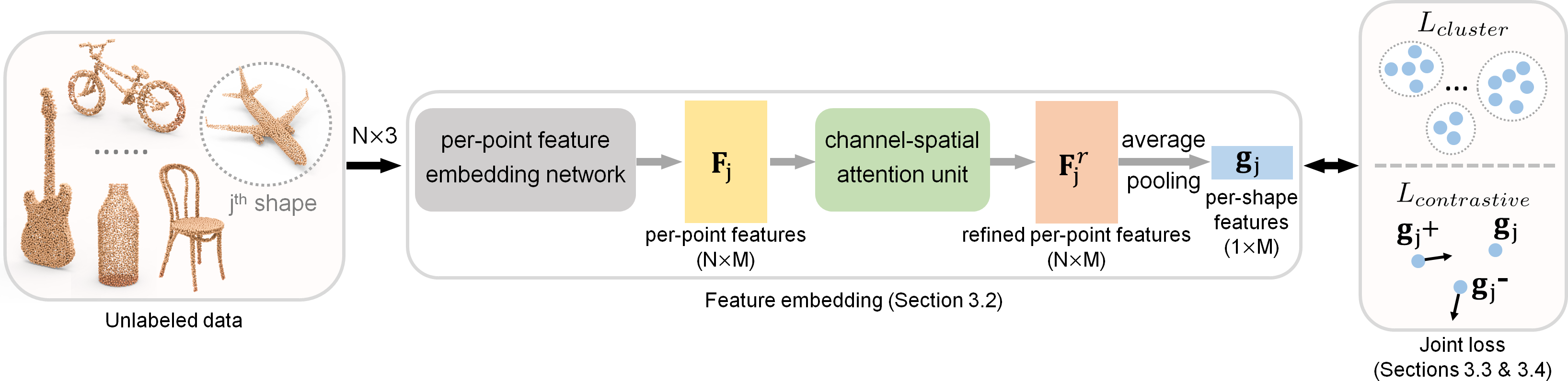}
	\vspace{-2mm}
	\caption{The overall framework of our unsupervised learning approach to detect distinctive regions on a given set of 3D shapes.}
	%\lqyu{Try to draw some details in ``per-point feature embedding" and reduce the size of ``per-shape feature embedding". Because the latter one is only a pooling operator.}}
	\label{fig:framework}
	%\vspace{-2mm}
\end{figure*}

%%%%%%%%%%%%%%%%%%%%%%%%%%%%%%%%%%%%%%%%%%%%%%%%%%%%%%%%%%%%

%\subsection{Saliency region detection}

\vspace*{-3pt}
\paragraph{Saliency region detection.}
%
%The notion of distinction is closely related to \emph{saliency} as they both measure regional importance.
%However, while distinction considers how unique a region is relative to objects of other types, saliency considers how unique and visible a region is relative to other regions within the same object.
Similar to distinction, saliency also measures regional importance of a shape, but it considers how unique and visible a region is relative to other regions {\em within the same object\/}. 
%\dc{just be aware that for saliency uniqueness is not enough...it must large enough - visible}
%
Lee~\etal~\shortcite{lee2005mesh} devised a scale-dependent measure to compute the mesh saliency, while Gal and Cohen-Or~\shortcite{gal2006salient} developed local surface descriptors to extract salient geometric features for partial shape matching and retrieval.
%Novatnack~\etal~\cite{novatnack2007scale} presented a salient region detection based on surface normals \dc{that is the key? the use of normals?}.
%%
These techniques are typically based on curvature, or other geometric features.
To alleviate the limitation of hand-crafted geometric features, several works adopt data-driven methods to effectively find the saliency for 3D surfaces~\cite{Chen:2012:SPO,shu2018detecting}.
In other aspects, Shtrom~\etal~\shortcite{shtrom2013saliency} detected saliency in large point sets, while Ponjou Tasse~\etal~\shortcite{ponjou2015cluster} detected saliency in point sets with a cluster-based approach.
Very recently, Wang~\etal~\shortcite{wang2018tracking} developed an eye tracking system to obtain mesh saliency from human viewing behavior.

%%%%%%%%%%%%%%%%%%%%%%%%%%%%%%%%%%%%%%%%%%%%%%%%%%%%%%%%%%%%

\vspace*{-3pt}
\paragraph{Network explanation.}
%\subsection{Network Explanation}
%
Our work is also related to the visualization of neuron activities in deep neural networks.
Zeiler and Fergus~\shortcite{zeiler2014visualizing} devised a perturbation-based method to find the contribution of each portion of the input by removing or masking them, and then running a forward pass on the new input to contrast with the original input. Such approach tends to be slow as the number of test regions grows.
Instead, backpropagation-based methods~\cite{shrikumar2017learning,sundararajan2017axiomatic,ancona2018towards,zhang2018top} compute the contribution of all the input regions in a single forward and backward pass through the network.
Unlike the backpropagation-based methods, Zhou~\etal~\shortcite{zhou2016learning} formulated the Class Activation Map (CAM) to localize the discriminative regions, and Selvaraju~\etal~\shortcite{Selvaraju2017grad} further developed the grad-CAM for producing visual explanations for decisions from the CNN models.
%Despite the promising visualization results, those methods are not unsupervised, indicating that class labels should be given for semi-supervised training.
%\dc{these methods are supervised??}
%These methods aims at visualizing the neuron activities of a pre-trained network, while in our work, we train a network to analyze and detect the distinctive regions of objects in a given data set.
These methods, however, require the class labels to visualize the neuron activities, while in our work, we analyze the network activities in an unsupervised training.

\vspace*{-3pt}
\paragraph{Deep neural networks for point set.}
%\subsection{Deep learning of point set}
%
%Rapid advances and demands in 3D sensing encourage the development of feature learning for point sets.
%PointNet~\cite{qi2016pointnet} and PointNet++~\cite{qi2017pointnet++} are pioneering works that directly consume point sets as input to neural networks.
%Successively, 
\NEW{Following PointNet~\cite{qi2016pointnet} and PointNet++~\cite{qi2017pointnet++} to embed features directly from point sets, 
several works successively introduce methods to improve the capturing of geometric information,~\eg, SpiderCNN~\cite{xu2018spidercnn}, KCNet~\cite{shen2018mining}, PointGrid~\cite{le2018pointgrid}, pointwise convolution~\cite{hua2018pointwise}, DGCNN~\cite{wang2018dynamic}, SPLATNet~\cite{su2018splatnet}, 
and PointCNN~\cite{li2018pointcnn}.
%
%TODO: to add later:
%PointWeb~\cite{}
%%
%Recently, Li~\etal~\shortcite{li2018pointcnn} presented PointCNN, which extends the notion of convolution from a local grid to an $\mathcal{X}$-convolution on a local set of points.
%The network encodes per-point features that are influenced by the points residing in their Euclidean neighborhood. 
Besides embedding point features for recognition tasks, several works propose to learn point features for registration,~\eg,~\cite{aoki2019pointnetlk,wang2019deep}.}
In our network, we adopt PointCNN as a module to extract point features, but other network models can also be used for the purpose.
%Note that our proposed method allows to plug in any state-of-the-art neural network freely.

%% file: method2.tex
\section{Method}
\label{sec:framework}

%%%%%%%%%%%%%%%%%%%%%%%%%%%%%%%%%%%%%%%%%%%%%%%%%%%%%%%%

\subsection{Overview}
\label{sec:problem_statement}
Given a set of shapes $\mathcal{S}=\{S_j\}_{j=1}^{N_\text{obj}}$,
%$\mathcal{S}=\{S_1, S_2, ..., S_M\}$, 
let $P_j = \{\mathbf{p}_{i,j}\}_{i=1}^{N}$ be a set of 3D points sampled on the $j$-th shape $S_j$, 
where 
$N_\text{obj}$ is the number of shapes in $\mathcal{S}$;
$N$ is the number of points in each point set $P_j$; and
$\mathbf{p}_{i,j}$\hspace*{1mm}$\in$\hspace*{1mm}$\mathbb{R}^3$ is the 3D coordinates of the $i$-th point in $P_j$.
The problem of detecting distinctive regions on shape $S_j$ is
\begin{quote}
To predict a per-point distinctiveness value $d_{i,j} \in [0, 1]$ for each point $\mathbf{p}_{i,j}$ on $S_j$ relative to the shapes in $\mathcal{S}$ that are of different types from $S_j$,
\end{quote}
%%We define the problem of detecting distinctive regions on shapes as predicting per-point distinctiveness $d_{i,j}$ for each point $p_{i,j}$ on $S_j$ relative to different shapes in $\mathcal{S}$.
\NEW{where a large $d_{i,j} \approx 1$ indicates that the associated region exists mainly in shapes of the same type as $S_j$ but not in shapes of other types, and a small $d_{i,j} \approx 0$ indicates that the associated region exists in all types of shapes.
Hence, $d_{i,j}$ indicates the degree to which the associated region distinguishes shape $S_j$ from others, and there is no requirement for the size of the distinctive regions.}

%\if 0
%\dc{something is missing in the problem statement.. the distinctiveness is relative to a given set.
%It is not independent to the set. 
%Given a set S, with $S_i$ objects...}
%\fi

Such goal requires us to consider not only the object itself, but also the other objects in the given reference set.
Intuitively, the designed network should contain per-point importance, while also having the ability to classify object types. 
Hence, we propose a novel framework to learn both per-point features and per-shape features, as shown in Figure~\ref{fig:framework}.
The per-point features are for calculating the distinctiveness $d_{i,j}$, while the per-shape features are for shape classification.
%We tackle this problem by learning the feature embedding for each point and then calculating the distinctiveness $d_{i,j}$ from the embedded features.
%%
To perform the feature embedding in an unsupervised manner, we drive the network to learn by solving an unsupervised shape clustering task \NEW{using our joint loss (see Figure~\ref{fig:framework} (right))}.

In the following, we first introduce the network for feature embedding (Section~\ref{subsec:feature_embedding}).
Next, we introduce a clustering-based nonparametric softmax classifier (Section~\ref{subsec:cluster_softmax}) and an adapted contrastive loss (Section~\ref{subsec:triplet}) to drive the unsupervised network to learn.
Lastly, we give details on the end-to-end network training (Section~\ref{subsec:end-to-end-training}), and describe how we obtain the per-point distinctiveness values from the embedded features (Section~\ref{sec:point_vis}). 
%%
%\lqyu{Actually, we want to do unsupervised feature embedding, the triplet loss is a  common and general manner to do that, however, this triplet loss is instance-level discrimination not cluster-level discrimination.}

%%%%%%%%%%%%%%%%%%%%%%%%%%%%%%%%%%%%%%%%%%%%%%%%%%%%%%%%

\subsection{Feature Embedding}
\label{subsec:feature_embedding}

\paragraph{Extracting the per-point features.} \
In this part, we aim to learn an embedding function $f_\theta$:
\begin{equation}
%\mathbf{f}_{i,j} = f_\theta(\mathbf{p}_{i,j}) \ ,
\NEW{\mathbf{F}_j = f_\theta(P_j) \ ,}
\end{equation}
\NEW{where $\mathbf{F}_j \in \mathbb{R}^{N \times M}$ is the set of extracted per-point features (see $\mathbf{F}_j$ in Figure~\ref{fig:framework}),
each row $\mathbf{f}_{i,j}$ in $\mathbf{F}_j$ is a per-point feature of $M$ channels, and
%which maps each point $\mathbf{p}_{i,j} \in P_j$ to feature $\mathbf{f}_{i,j} \in \mathbf{F}_j$.
$f_\theta$ is a deep neural network with parameters $\theta$.
%to map each point $\mathbf{p}_{i,j}$ to feature $\mathbf{f}_{i,j}$}.
%
In theory, 
%the embedded per-point feature 
$\mathbf{f}_{i,j}$ should represent the underlying local geometric structures around each point and further reveal the point's distinctiveness.
%We apply a point-set-network to extract per-point feature $\mathbf{f}_{i,j}$ with $M$ channels, and the per-point features of the same object can be stacked together as the embedded feature $\mathbf{F}_j$ for the $j$-th object;
We apply a point-set-network to extract per-point features, and the choice of the point-set-network is flexible.}
Most recent networks on processing point sets can be employed; here, we adopt the segmentation architecture of PointCNN~\cite{li2018pointcnn} to learn $f_\theta$.
%Since the purpose of learning $f_\theta$ is not for per-point predicton but for point distinction detection, only applying a pure segmentation network may not well endow each point with distinctive information. Considering that the attention mechanism could drive the network to pay more attention to important points by attending to all other points, we thus further introduce a channel-spatial attention unit~\cite{woo2018cbam} to tell the network where to focus and also improve the representation of interests.

\begin{figure}[!t]
	\centering
	\includegraphics[width=1.0\linewidth]{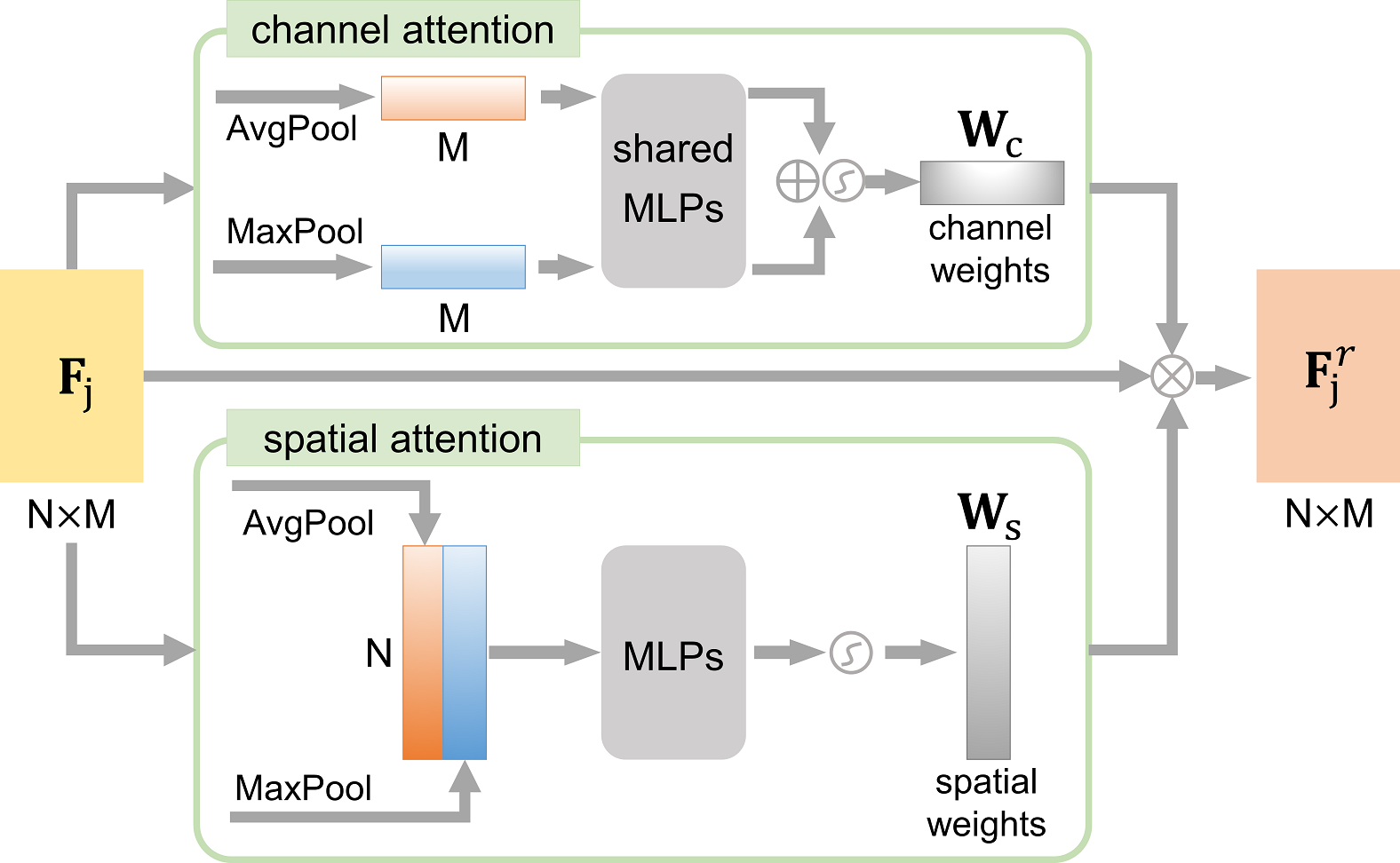}
	\caption{\phil{The channel-spatial attention unit}.}
	\label{fig:attention}
	%\vspace*{-2mm}
\end{figure}

%\phil{\paragraph{Re-calibrating the per-point features.} \ 
\paragraph{\NEW{Adaptive per-point features refinement.}} \ 
So far, the per-point features are extracted locally over each given shape by the point-set network.
As the distinctiveness requires shape-level context, not simply local context around each point, we further propose to \NEW{refine} the per-point features $\mathbf{F}_{j}$.
To this end, we adopt~\cite{woo2018cbam} to formulate the channel-spatial attention unit (see Figure~\ref{fig:attention}) to fuse the $M$ feature channels over the $N$ per-point features together and produce the \NEW{refined} per-point embedding feature $\mathbf{F}^r_j$.

%a per-shape global feature $\mathbf{g}_j$ for each point set $P_j$.
%%
%Simply stacking, or concatenating, the per-point features together, i.e., $\mathbf{F}_j$, may not produce the best per-shape features for predicting the point distinctiveness on the shapes.
%To produce the best per-shape features for predicting the point distinctiveness on the shapes, we first extend~\cite{woo2018cbam} to formulate the channel-spatial attention unit in our network (see Figure~\ref{fig:framework}), to fuse the $M$ feature channels and the $N$ per-point features together for producing the overall re-calibrated per-point embedding feature $\mathbf{F}^r_j$.
%Next, we obtain the per-shape global feature $\mathbf{g}_j$ from $\mathbf{F}^r_j$.
%
%Considering that the attention mechanism could dig the relationship of $N$ different points and $M$ different features by attending to all other points and features, we thus incorporate a channel-spatial attention unit~\cite{woo2018cbam} into our network to tell the network where to focus and also improve the representation of interests.

%%\phil{As shown in Figure~\ref{fig:attention}, the architecture of our employed channel-spatial attention unit has a channel attention sub-unit and a %%spatial attention sub-unit.
%%The purpose of the channel attention sub-unit is to exploit the inter-channel relationship in the features, while the spatial attention sub-unit is to %%exploit the inter-spatial (or inter-points) relationship in the features.
%%%%
%%Here, we first aggregate the spatial information of $\mathbf{F}_j$ by using both average-pooling and max-pooling operations, and forward the two %%aggregated features to a series of shared MLPs.
%%Then, we can obtain the channel weights $\mathbf{W}_c \in \mathbb{R}^{1\times M}$ by merging the two output feature vectors with summation operation and %%sigmoid function.
%%%%
%%Similarly, for the spatial attention, we first aggregate over the channels in $\mathbf{F}_j$ by using both average-pooling and max-pooling.
%%We then concatenate the two features and feed them to MLPs.
%%The spatial weights $\mathbf{W}_s \in \mathbb{R}^{N\times 1}$ are obtained by applying a sigmoid function to the output feature vector from the MLPs.
%%Lastly, we obtain $\mathbf{F}^r_j$ by multiplying $\mathbf{W}_c$ and $\mathbf{W}_s$ to $\mathbf{F}_j$ in both channel-wise and spatial-wise manner.}
%%

\paragraph{Extracting the per-shape feature.} \ 
Further, we use an average-pooling operation to obtain the global feature $\mathbf{g}_j$ from $\mathbf{F}^r_j$:
%%To obtain the per-shape global feature $\mathbf{g}_j$,
%%we further use an average pooling on $\mathbf{F}^r_j$ to obtain the global feature $\mathbf{g}_j$:
%%
\begin{equation}
\label{equ:pooling}
\mathbf{g}_j = {\frac{\sum_{i=1}^{N}\mathbf{f}^r_{i,j}}{N}},
\end{equation} 
where $\mathbf{f}^r_{i,j}$ denotes the \NEW{$i$-th per-point feature vector in $\mathbf{F}^r_j$}.
%after \NEW{refinement}.
As shown in Figure~\ref{fig:framework}, the process of training the network to learn the local (per-point) and global (per-shape) features is driven by an unsupervised joint loss function, which we shall elaborate below.

\subsection{Clustering-based Nonparametric Softmax}
\label{subsec:cluster_softmax}
To drive the network to classify objects in an unsupervised manner, we propose a clustering-based nonparametric softmax classifier.
In a typical supervised deep neural network for classification, the softmax classifier is commonly employed and the probability of the $j$-th object being recognized as the $q$-th class is
\begin{equation}
\label{equ:softmax}
P(y_j=q|\mathbf{g}_j)=\frac{exp(\mathbf{w}_q^T\mathbf{g}_j)}{\sum_{k=1}^{C}exp(\mathbf{w}_k^T\mathbf{g}_j)} \ ,
\end{equation}
where
%$\mathbf{g}_j$ is the global feature of the $j$-th object,
$y_j$ is the class label of the $j$-th object;
$C$ is a hyperparameter that denotes the number of classes;
$q \in \{1,...,C\}$ is the class assignment;
%for the $j$-th object;
$\mathbf{w}_k$ is the weight vector for the $k$-th class;
$k \in \{1,...,C\}$; and 
$\mathbf{w}_k^T\mathbf{g}_j$ measures how well $\mathbf{g}_j$ matches the $k$-th class, so $\mathbf{w}_k$ serves as the class prototype of the $k$-th class.
%say for class $k$.\lqyu{too many say and say for...}

In supervised learning, we can leverage the class labels provided in training data to learn the class prototype $\mathbf{w}_k$ of each class.
This is, however, not possible for unsupervised learning.
%In unsupervised learning, $\mathbf{w}_k$, however, cannot be learned, since there are no class labels for supervision.
%
Recently, an observation was reported by Liu~\etal~\shortcite{liu2018transductive} that when the network has successfully converged, the class prototype is usually consistent with the average of all the feature vectors belonging to the same class.
\NEW{Based on this observation, we thus approximate the class prototype by using the average of all the feature vectors belonging to the class, since no class labels are given in our setting.}
%This tells us that the class prototype does not have to be learned by the class labels, but 
%can be approximated by the average of all the feature vectors belonging to the associated class, if no class labels are given.}

%\phil{
%In the case of our unsupervised setting, we have {\em clusters\/} but not classes, since we do not even have the class labels.
%\phil{what does it mean by direction of the data centroid?}
%
%, the average feature vector within a class can serve as the class prototype.
We adopt the above observation to our problem by formulating the \emph{clustering-based nonparametric softmax classifier}, where we iteratively re-cluster the per-shape feature vectors $\mathbf{g}_j$ in the network and take the average feature vector of each cluster to estimate the cluster prototype $\mathbf{w}_k$.
%\footnote{Here we regard different clusters as different classes.}.
%the class prototypes (i.e., $\mathbf{w}_k$).
% and yet, we do not explicitly create any class label.
In this way, we can approximate the probability $P(y_j=q|\mathbf{g}_j)$ for unsupervised learning as
\begin{equation}
\label{equ:cluster_softmax}
P(y_j=q|\mathbf{g}_j)
\approx
\frac{exp(\bar{\mathbf{g}}_q^T\mathbf{g}_j/\tau)}{\sum_{k=1}^{C}exp(\bar{\mathbf{g}}_k^T\mathbf{g}_j/\tau)},
\end{equation}
where
$\bar{\mathbf{g}}_k$$=$$\frac{1}{|\mathbb{C}_k|}\sum_{t\in{\mathbb{C}_k}}\mathbf{g}_t$ is the average feature vector over all per-shape global feature vectors $\mathbf{g}_t$ of cluster $\mathbb{C}_k$; 
we take $\bar{\mathbf{g}}_k$ (per-cluster) to approximate $\mathbf{w}_k$ for unsupervised learning; and
$C$ denotes the number of clusters in our unsupervised setting.
Further, we enforce $\Arrowvert \mathbf{g}_j \Arrowvert$$=$$1$ via an L$_2$-normalization layer in the network and make use of $\tau$, which is a temperature parameter, to control the concentration level of the distribution~\cite{hinton2015distilling,wu2018unsupervised}.

In our experiments, we set $\tau$ as 0.07, following the setting in~\cite{wu2018unsupervised}.
Then, our learning objective is to maximize $P(y_j$\hspace*{1mm}$=$\hspace*{1mm}$q|\mathbf{g}_j)$, or equivalently, to minimize the negative log-likelihood of the probability.
%, where $q_j$ is the class id of $\mathbf{g}_j$.
%\lqyu{I think the reviewer may have questions about what `q' is.}
Therefore, our {\em clustering-based nonparametric softmax loss\/} is formulated as
\begin{equation}
\label{equ:cluster_loss}
L_{cluster}=-\sum_{j=1}^{N_\text{obj}}\log P(y_j=q|\mathbf{g}_j).
\end{equation}
%where $q$ is the cluster assignment for shape $S_j$ after the clustering procedure.
%%
%We need to pre-determine the class id $q_j$ of each $\mathbf{g}_j$ to calculate the above probability.
In our implementation, we use spectral clustering~\cite{stella2003multiclass,von2007tutorial} to cluster the per-shape global features $\mathbf{g}_j$ 
%into $C$ clusters 
in each training epoch.
% and then compute $\bar{\mathbf{g}}_k$ of each cluster.
Experimentally, we found that our network detects similar distinctive regions when equipped with different clustering algorithms; see Supplementary Material Part A for the evaluation.
\phil{Also, please see Supplementary Material Part B for the effect of having different $C$ on extracting distinctive regions.}
%\phil{I suggest that it is better not to say too much here}
%, by which we can control a contour-to-detail learning of distinctive regions on a given set of shapes.}

%\xz{In Supplementary Material Part B, we will elaborate on the effect of having different $C$, by which we can control a contour-to-detail learning of distinctive regions on a given set of shapes.}
%In Section~\ref{subsec:cluster_number}, we will elaborate on the effect of having different $C$, by which we can control a coarse-to-fine learning of distinctive regions on a given set of shapes.

%\lqyu{In another ECCV2018 paper, they discuss how to avoid the cluster generating trivial solutions, such as empty clusters. So do we have the same issues?}
%%
%%
\if 0
\lqyu{Note that it is flexible to set the different $C$ in the clustering. We can control the distinctive area by easily adjusting cluster number $C$, meaning that it allows us to choose coarser or finer differences in the database.
	In the extreme setting with $C=M$, we can detect the distinctive regions of an individual against the rest of the set shapes.}
\phil{we may say these later in sec 4}
\fi
%%

%%%%%%%%%%%%%%%%%%%%%%%%%%%%%%%%%%%%%%%%%%%%%%%%%%%%%%%%

\subsection{Adapted Contrastive Learning}
\label{subsec:triplet}
Inaccurate clustering results are inevitable, so relying only on the clustering-based loss may mislead the learning of the network.
\NEW{Motivated by~\cite{hjelm2018learning,henaff2019data,bachman2019learning},} to stabilize and enhance the feature learning in the network, we formulate an {\em adapted contrastive loss\/}, which is particularly important at the beginning of the training process when the clustering results are more random.
Considering input point set $P_j$ to the network as the \emph{anchor}, for each training epoch, we form a {\em positive\/} point set sample $P_j^+$ and a {\em negative\/} point set sample $P_j^-$ for $P_j$, such that the per-shape global feature $\mathbf{g}_j^+$ associated with $P_j^+$ is close to $\mathbf{g}_j$, while the per-shape global feature $\mathbf{g}_j^-$ associated with $P_j^-$ is far from $\mathbf{g}_j$.
\begin{itemize}
%
%\vspace*{-1mm}
\item
%For $P_j^-$, we assign to $\mathbf{g}_j^-$ a global feature randomly picked from the clusters that $P_j$ does not belong to.
For $P_j^-$, we randomly pick a point set from the shapes in the clusters that $P_j$ does not belong to.
%
%\vspace*{-1mm}
\item
For $P_j^+$, since the intra-class clustering results may not be reliable, especially at the beginning of the training process, we thus do not randomly pick from the cluster that $P_j$ belongs to.
Rather, we resample another point set $P_j^+$ on the given 3D shape ($S_j$) associated with $P_j$ and pass $P_j^+$ to the network to generate $\mathbf{g}_j^+$.
%\phil{
Note that $P_j^+$ and $P_j$ are different point sets due to randomness in the point sampling process, but essentially, they describe the same object,~\ie, $S_j$.
%Hence, both $P_j^+$ and $P_j$ describe the same object $S_j$ but with different point distribution.
%We found that this strategy can stabilize the training process and help the model to converge.
%to accelerate the convergence.\phil{I down-tone a bit}
%
\end{itemize}
%
%We found that this strategy can stabilize the training process and help the model to converge.
%\lqyu{maybe we can only describe how we do, not add argument.}
%\phil{TODO: XZ - really? you should read the comments from LQ and decide if we can claim or not; if not, comment it out as LQ suggested... else put it back}
%
%\vspace*{-1mm}
%Given that $\mathbf{g}_j^+$ should be very close to $\mathbf{g}_j$ due to the above sampling strategy, 
We take the above triplet \{$\mathbf{g}_j$, $\mathbf{g}_j^+$, $\mathbf{g}_j^-$\} to form an {\em adapted contrastive loss\/} following~\cite{hadsell2006dimensionality} as
\begin{equation}
\label{equ:triplet_loss}
L_{contrastive}=D(\mathbf{g}_j, \mathbf{g}_j^+) + \max (0, \lambda - D(\mathbf{g}_j, \mathbf{g}_j^-)),
\end{equation}
where $D$ is the Euclidean distance in feature space, and we set $\lambda=2.0$ in our experiments.
Importantly, we generate such triplet input {\em dynamically\/} for each $P_j$ in each training epoch.
Using this strategy, we can increase the diversity of the training samples and produce more reliable samples as the training progresses.

%\lqyu{what is the value of $\lambda$ in other paper? It seems that $D(\mathbf{g}_j, \mathbf{g}_j^-)$ always less than 2.}
%\phil{TODO: I haven't checked this argument... XZ?}

%%%%%%%%%%%%%%%%%%%%%%%%%%%%%%%%%%%%%%%%%%%%%%%%%%%%%%%%

\subsection{End-to-end Network Training}
\label{subsec:end-to-end-training}
Overall, we end-to-end train the network to learn the features for clustering the given shapes using the joint loss function
\begin{equation}
\label{equ:totalloss}
L(\theta) = L_{cluster} + \alpha L_{contrastive} + \beta \Arrowvert\theta\Arrowvert^2,
\end{equation}
where $\alpha$ balances the two loss terms and $\beta$ is the multiplier of weight decay in the regularization term \NEW{(see Section~\ref{ssec:imp_details} for their values)}.
%%
%In the first training epoch, since the per-shape features $\mathbf{g}_j$ are randomly set, we choose to apply the cluster loss starting from the second epoch.
%\phil{I'm not sure if I catch what you meant...}\xz{No, it is wrong, we do the clustering in the initial epoch. That's why the clustering result at the beginning is not reliable.}

%%%%%%%%%%%%%%%%%%%%%%%%%%%%%%%%%%%%%

\begin{figure}[!t]
	\centering
	\includegraphics[width=1.0\linewidth]{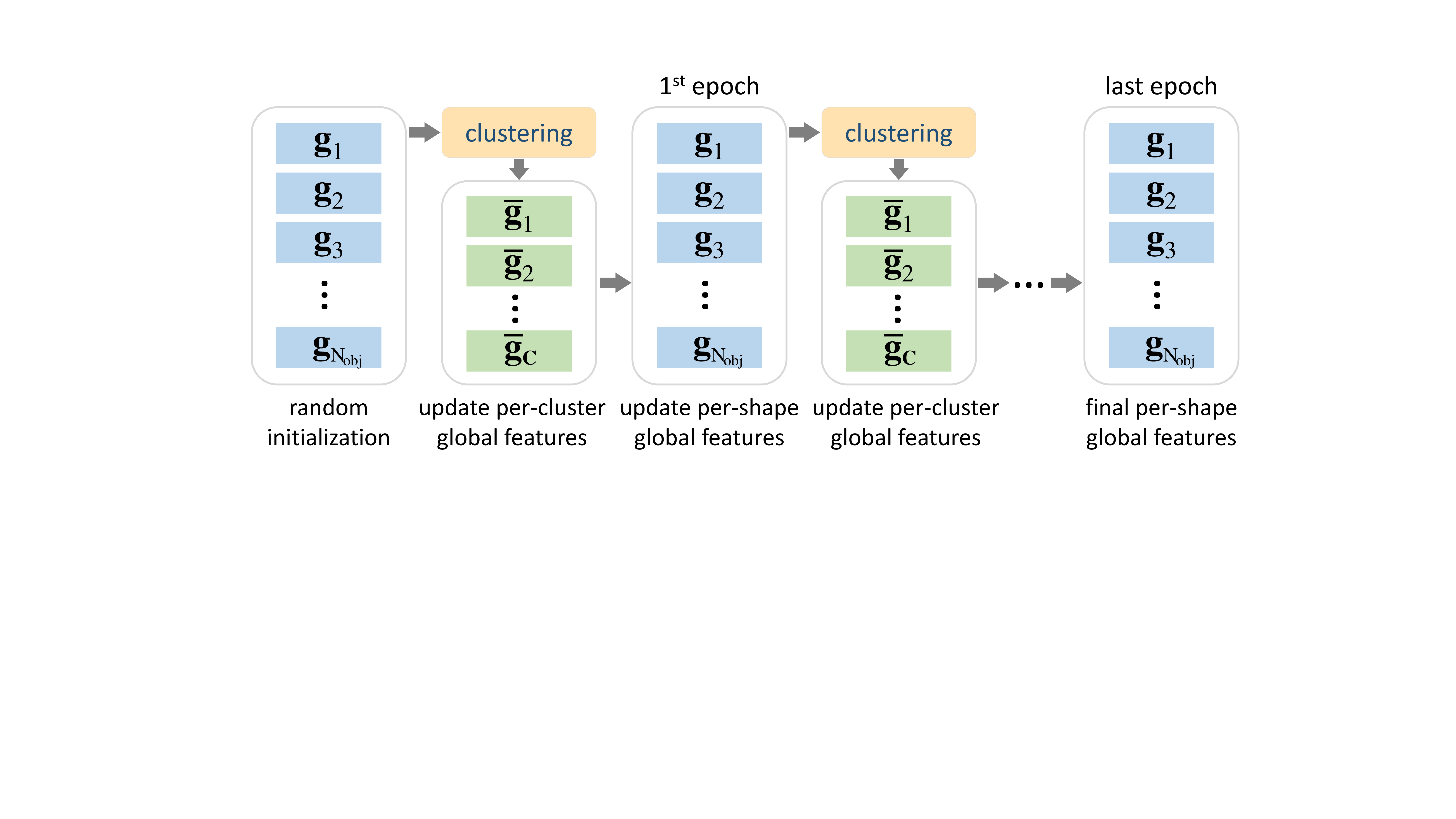}
	\caption{Illustration of the network training process.}
	\label{fig:training_process}
	%\vspace*{-2mm}
\end{figure}

%\vspace*{2mm}
In summary, the feature embedding is conducted in a self-training way by iteratively learning the feature vectors, re-clustering them,
%\NEW{into $C$ clusters}, %\phil{maybe too much to say this again and again}
then using the clustering results to fine-tune the model.
% until the model converges.
%
Figure~\ref{fig:training_process} illustrates the whole training process, where we first randomly initialize the per-shape global feature $\mathbf{g}_j$ of each training sample $P_j$, \NEW{cluster them into $C$ classes,} and generate the per-cluster global feature $\bar{\mathbf{g}}_k$ for each cluster.
Early in the training, these $\mathbf{g}_j$ and $\bar{\mathbf{g}}_k$ are unlikely reliable, but as the training progresses, we iteratively update these per-shape and per-cluster features in each training epoch, these features can then gradually converge and let us further obtain the per-point distinctiveness in the given shapes.
\NEW{Here, $C$ remains unchanged during the training, and our method does not require even class size in the training data.
The clustering method we employed,~\ie, spectral clustering, will divide the training samples automatically into $C$ clusters based on the feature similarity.}
%
%Figure~\ref{fig:tsne} presents t-SNE visualizations~\cite{maaten2008visualizing} of the global per-shape features ($\mathbf{g}_j$) during the unsupervised training process, where we iteratively cluster over 9,000 shapes into 40 classes.
%, indicating that the features become discriminative as the training process.
%\dc{How it indicates that ? these are the per-shape features!!! not the the discriminative per-point features. Also we do not need to explain what tsne is. Let me suggest the following text}
%%%%%%

\begin{figure}[!t]
	\centering
	\includegraphics[width=0.9\linewidth]{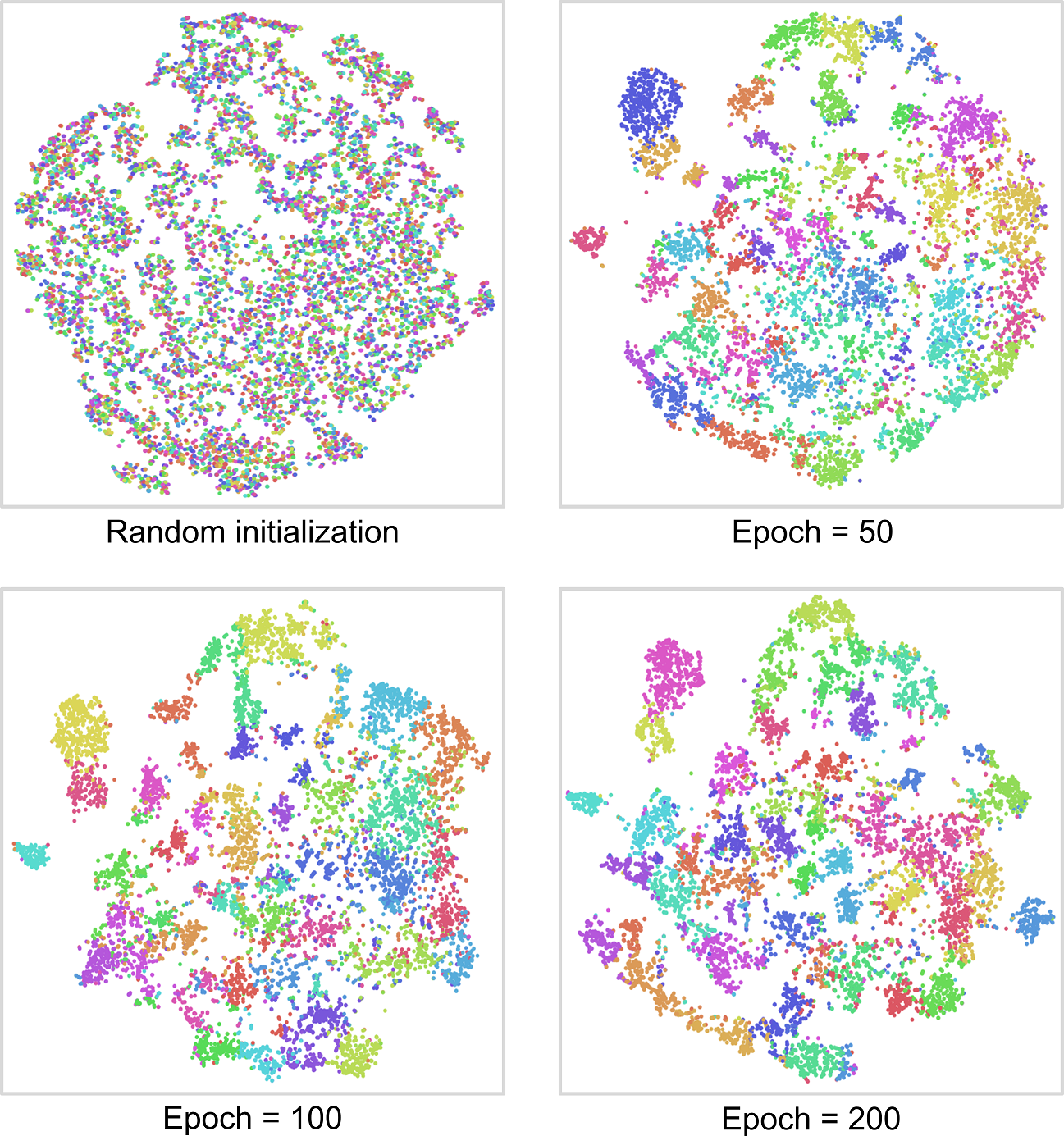}
	\caption{t-SNE visualizations of the per-shape features clustering during the unsupervised training process.}
	\label{fig:tsne}
	%\vspace*{-2mm}
\end{figure}

Figure~\ref{fig:tsne} shows t-SNE visualizations that reveal the clustering of the per-shape features ($\mathbf{g}_j$) during the unsupervised training.
Here, we cluster over 9000 shapes into 40 classes.
The dimension of the features (\ie, $M$) is 128 in our implementation.

\subsection{Obtaining and Visualizing the Distinctiveness}
\label{sec:point_vis}

As introduced earlier in Section~\ref{sec:problem_statement}, we design our unsupervised approach to learn both per-shape and per-point features in the given shapes.
After the training to meet the shape clustering task, the response of the activation neuron associated with the per-point features should positively correlate to its confidence of the detection~\cite{zhang2018top}.
Therefore, we obtain the per-point distinctiveness $d_{i,j}$ from $\mathbf{f}^r_{i,j}$ of each point $\mathbf{p}_{i,j}$ by taking the maximum value in $\mathbf{f}^r_{i,j}$ and normalizing $d_{i,j}$ between 0 and 1 for each shape.
For a comparison of applying other alternatives to extract $d_{i,j}$ from $\mathbf{f}^r_{i,j}$, including the mean, $L_2$ norm, average of the three largest values, etc., please refer to Supplementary Material Part C.

\begin{figure*}[htbp]
	\centering
	\includegraphics[width=0.95\linewidth]{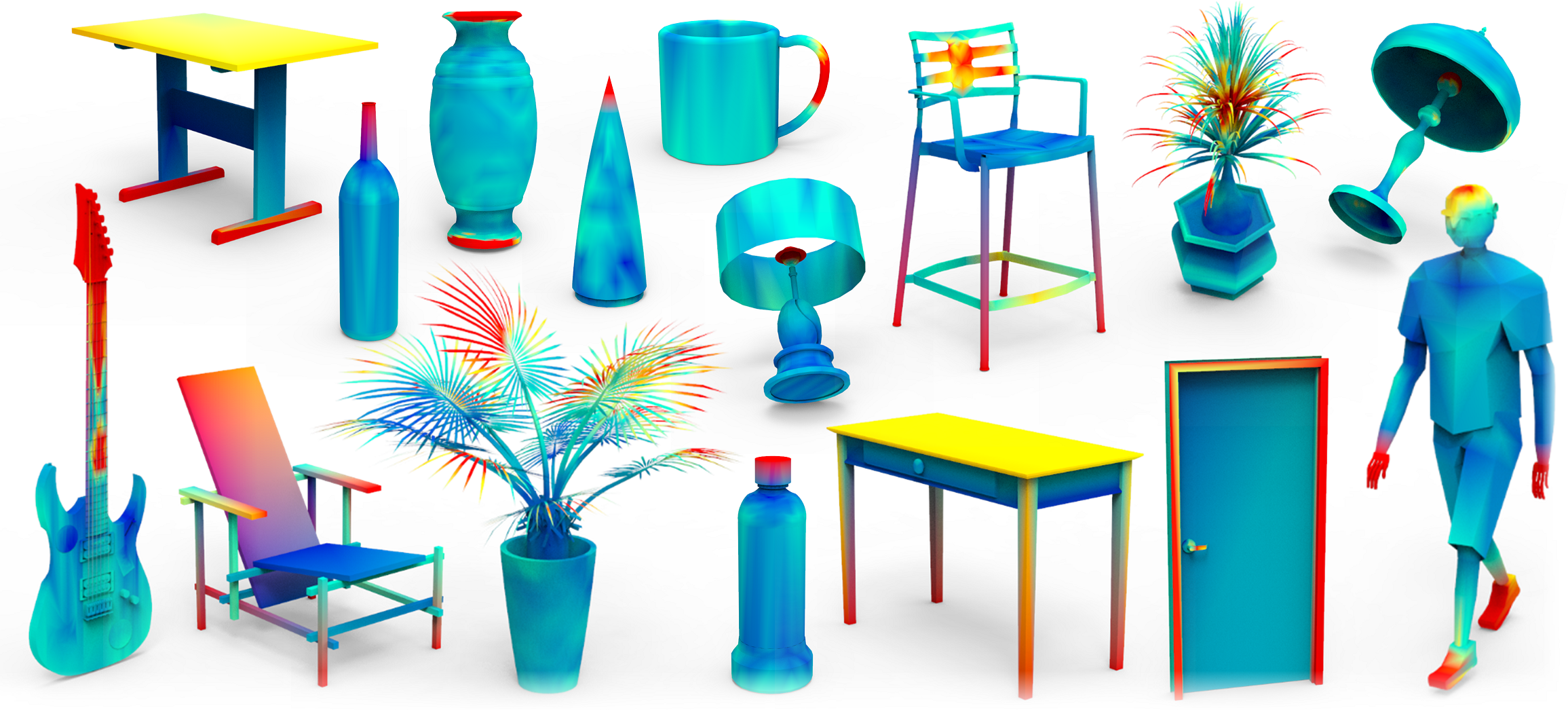}
	\vspace*{-2mm} % due to soft shadow below the image
	\caption{Distinctive regions detected by our method \phil{\em unsupervisedly\/} on various 3D models in ModelNet40~\cite{wu20153d}; red indicates high distinctive regions.}
	\label{fig:vis}
	\vspace*{-1mm}
\end{figure*}

Furthermore, to visualize the distinctiveness results, we project the per-point distinctiveness on point set $P_j$ back to the original shape $S_j$ and obtain a distinctiveness value for every vertex on the original shape by averaging the distinctiveness values over the nearby sampled points in $P_j$; see Figure~\ref{fig:teaser1} and Figure~\ref{fig:vis} for example results, where regions in red are the most distinctive.

%%%%%%%%%%%%%%%%%%%%%%%%%%%%%%%%%%%%%

%%%%%%%%%%%%%%%%%%%%%%%%%%%%%%%%%%%%%%%%%%%%%%%%%%%%%%%%

%% file: experiment.tex
%%%%%%%%%%%%%%%%%%%%%%%%%%%%%%%%%%%%%%%%%%%%%%%%%%%%%%%%

\section{Experiments and Results}
\label{sec:experiment}

%%%%%%%%%%%%%%%%%%%%%%%%%%%%%%%%%%%%%%%%%%

\subsection{Implementation Details}
\label{ssec:imp_details}

Our method was implemented using TensorFlow~\cite{45381}.
To train the network, we randomly sampled 2,048 points for each shape as input and augmented the input point sets on-the-fly, including random rotation, scaling, shifting, and jittering.
Moreover, we empirically set $\alpha$ and $\beta$ in Eq.~\eqref{equ:totalloss} as $3.0$ and $10^{-5}$, respectively, and trained our network for 200 epochs using the Adam optimizer~\cite{kingma2014adam} with a learning rate of 0.01.
\NEW{See Supplementary Material Part J for further analysis on $\alpha$ and $\beta$.}
%\lqyu{why setting as 3, and the batch size.}
%{\footnotesize
%\color[rgb]{0.2,0.8,0.2}
%\begin{verbatim}
%\phil{note:} use \verb|Eq.~\eqref| instead of \verb|Equ.~\ref|
%\end{verbatim}
%}
%
\NEW{For the ModelNet40 dataset~\cite{wu20153d} with 9,839 training samples, it took about 25 hours to train our network on a 12GB TITAN Xp GPU with a batch size of 50. The network has 0.48M parameters. For inference, it took about 0.017 seconds for our method to predict the per-point distinctiveness values for a point cloud of 2,048 points.}

Based on PointCNN~\cite{li2018pointcnn}, we further made the following adaptations to the feature embedding component in the network architecture.
First, we use a fixed-size query ball~\cite{qi2017pointnet++} instead of KNN or geodesic-like KNN~\cite{yu-2018-EC-net} to find the local neighborhood for extracting the point features; \NEW{note also that according to~\cite{qi2017pointnet++}, query ball
%~\cite{qi2017pointnet++} 
is preferred for finding the local neighborhood in tasks that require local pattern recognition, e.g., our distinctive detection task.}
%\NEW{since for distinctive detection, we need local pattern recognition to predict per-point distinctiveness.}
%Also, we found that the query ball is more effective for our problem, since the extracted local structures have fixed spatial size.
%%
Second, we removed the $\mathcal{X}$-conv operation in the deconvolution part of the PointCNN segmentation network and directly used feature interpolation~\cite{qi2017pointnet++} for per-point feature restoration.
In this way, we can reduce the number of network parameters and speed up the network training process with little degradation in the quality of the results.
Lastly, we explored different network backbones (\ie, PointNet and PointNet++) for learning the per-point local features; see Supplementary Material Part D for the experimental results.

\subsection{Detecting Distinctive Regions}
\label{subsec:distinctive}

%%%%%%%%%%%%%%%%%%%%%%%%%%%%%%%%

\paragraph{Distinctiveness visualization.} \
We employed the ModelNet40 training split dataset~\cite{wu20153d} and trained our network in an unsupervised manner to sort the models in the dataset into $40$ clusters.
Figure~\ref{fig:vis} shows the distinctive regions detected by our method on a variety of models in the dataset, where red color indicates high distinctive regions.
When we determine if a region is distinctive, \NEWNEW{instead of looking just at the shape itself, we consider all shapes of different classes (but not shapes of its own class) in the dataset.}
%\NEW{When we determine if a region is distinctive, we \NEWNEW{only consider (all) shapes of other different} classes in the dataset, instead of just looking at each individual shape.}
%
Taking the Person shape in Figure~\ref{fig:vis} (right) as an example, the head, feet, and hand are found to be more distinctive (red), while the body part is less (blue).
Since we compare the Person shapes with shapes of other types, these human parts are distinctive for the network to recognize the Person shapes relative to others.
\NEW{For the two Lamp shapes in Figure~\ref{fig:vis} (middle \& top-right), our method detects the bulb as distinctive, since it is common and unique in this class compared with shapes in other classes.
%Note that, since not all the lamps have a lampshade, our method does not detect lampshades as distinctive.
For the Chair shapes, not only the legs are detected, the back is also detected as distinctive.}
Similarly, our method detects as distinctive the handle of the Cup, the leaves of the Plant, the struts of the Guitar, the handle of the Door, etc.
\NEW{For more results, please refer to Supplementary Material Part K.}

\begin{figure}[!t]
	\centering
	\includegraphics[width=0.96\linewidth]{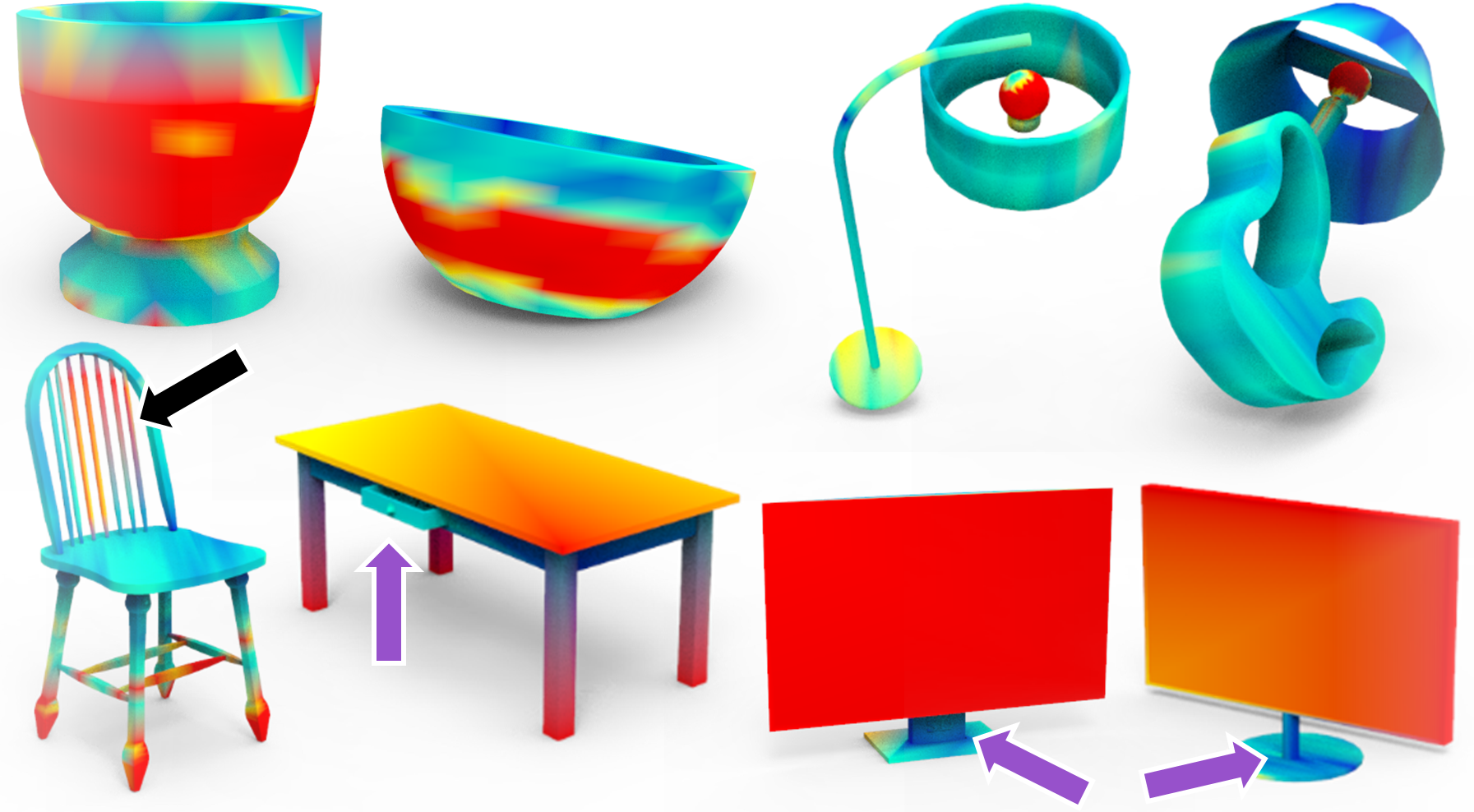}
	\vspace*{-2mm}
	\caption{\NEW{Distinctive regions detected by our method may not simply lie on high-curvature and extreme regions, as shown in these examples.
	Also, not all extreme regions are detected as distinctive (see purple arrows).}}
	\label{fig:not_extreme}
%	\vspace*{-1.5mm}
\end{figure}

\NEW{\NEWNEW{In particular}, our network does not simply locate high-curvature regions and extremities as distinctive.
We show several another examples in Figure~\ref{fig:not_extreme}, where the detected distinctive regions are not extreme regions.
%Besides the back of the two Chair shapes in Figure~\ref{fig:vis}, we show several another examples in Figure~\ref{fig:not_extreme}.
%For the shapes shown in the figure, the detected distinctive regions are not extreme regions.
%For the four shapes in the top row, the detected distinctive regions are not extreme regions at all.
%For the four shapes in the bottom row, the detected distinctive regions are not only extreme regions.
For example, for the four shapes shown in the top row, the detected distinctive regions are not extreme regions at all.
Concerning the other four shapes on the bottom, for the Chair shape, its back (see the region marked by the black arrow), which is not extreme region, is also detected.
%for the Chair shape, the non-extreme back region is also detected. 
For the Table and Monitor shapes, some obvious extreme regions are not detected (see the regions marked by the purple arrows).
For more non-extreme examples, please see Supplementary Material Part M.}

%%%%%%%%%%%%%%%%%%%%%%%%%%%%%%%%
%%%%%%%%%%%%%%%%%%%%%%%%%%%%%%%%

\begin{figure*}[t]
	\centering
	\includegraphics[width=0.945\linewidth]{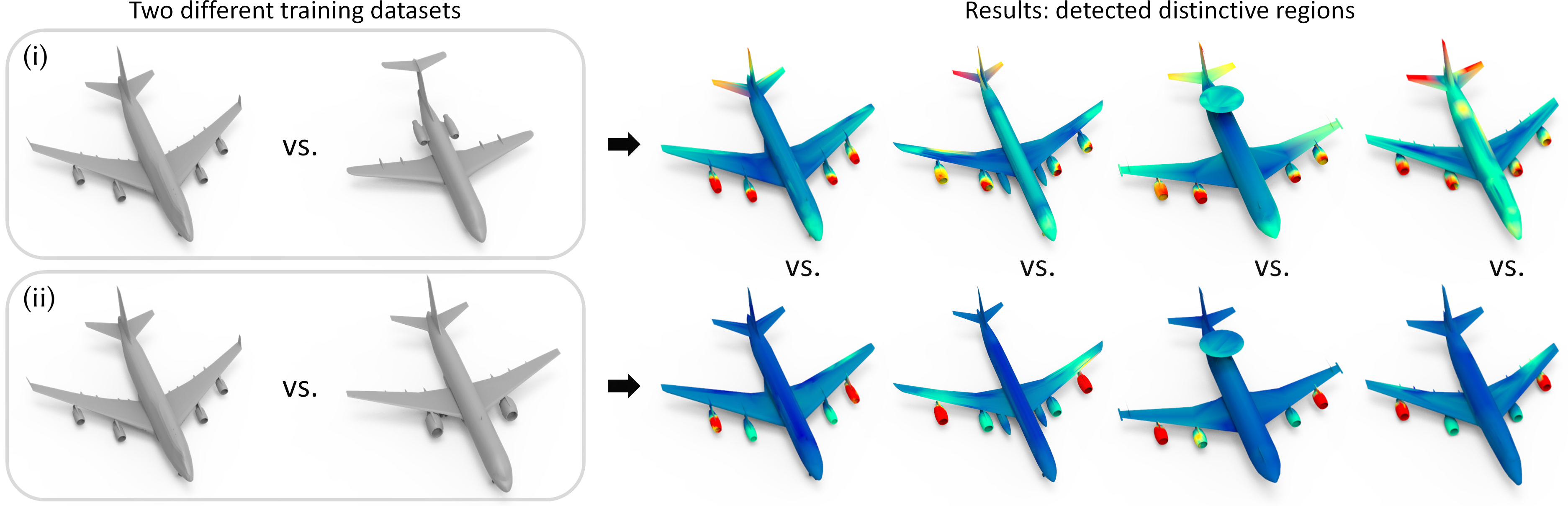}
	\vspace*{-2mm}
	\caption{Our network, when trained with different datasets (left), detects different distinctive regions on the same given objects (right): (i) a training set of four-engine and tail-engine airplanes in the top row, and (ii) a training set of four-engine and two-engine airplanes in the bottom row.}
	\label{fig:database}
	%\vspace*{-1mm}
\end{figure*}

\begin{figure}[!t]
	\centering
	%\vspace*{-2.5mm}
	\includegraphics[width=0.9\linewidth]{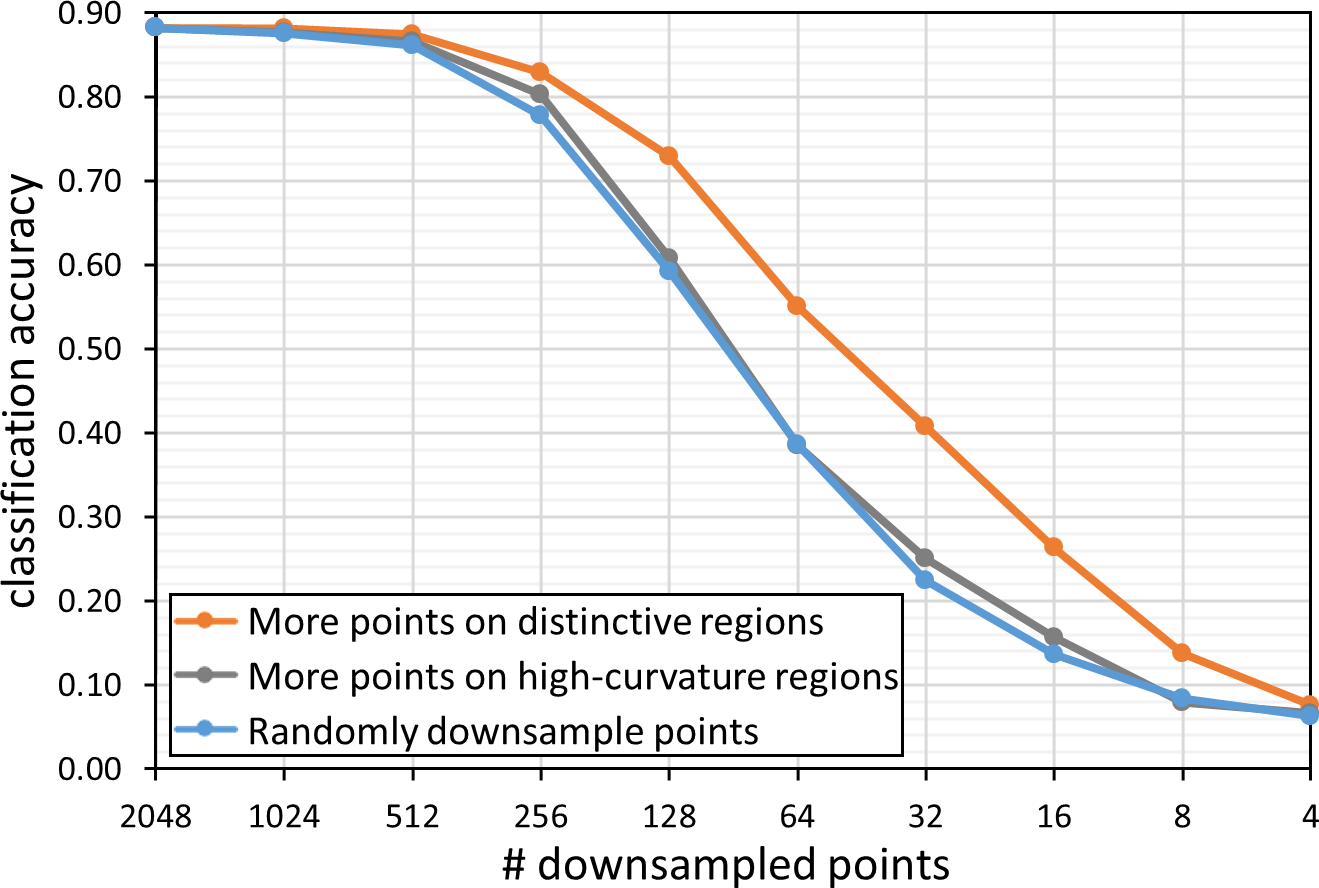}
	%\vspace*{-2mm}
	\caption{Overall shape classification accuracy 
	%on ModelNet40 
	when we downsample points on test shapes using three different preferences.
	\NEW{Having more distinctive points detected by our method (orange plot) leads to higher accuracy.}}
	%: downsample points more or less on distinctive regions detected by our method.}
	%
	%	\caption{Changes in the overall shape classification accuracy when we sub-sample the testing point sets (from ModelNet40) using different strategies: keeping more distinctive points (blue plot), random dropping (gray plot), and dropping more distinctive points (orange plot).
	%		Comparing the three plots, we can see the impact of the distinctive points detected by our method on the shape classification accuracy.\phil{TODO}}
	\label{fig:distinctive_test}
	%	\vspace*{-2.5mm}
\end{figure}

\vspace*{-3pt}
\paragraph{Quantitative evaluation.} \
Next, we quantitatively evaluated {\em how helpful the detected distinctive regions are to shape classification\/}.
Here, we employed totally 2,468 models in the ModelNet40 testing dataset as test shapes, and used three different preferences to downsample points from pre-sampled point sets ($N=2,048$) on the test shapes:
(i) probability to preserve a point based on the distinctiveness at the point; 
(ii) probability to preserve a point based on the curvature at the point; and
(iii) ignore the point importance and downsample points at random.
Therefore, preference (i) leads to the production of more points on our detected distinctive regions, while preference (ii) leads to the production of more points on high-curvature regions.
Then, we fed the downsampled points into a classification network (\ie, PointNet~\cite{qi2016pointnet}), which has been pre-trained on the ModelNet40 training split dataset with 2,048 points on each shape for a shape classification task, and computed the overall classification accuracy averaged over all the test shapes.
%Note that the PointNet++ is trained using the ModelNet40 training dataset and can take different number of points as input.

Figure~\ref{fig:distinctive_test} plots the overall classification accuracy for point sets downsampled with the three different preferences using decreasing number of downsampled points.
%As an example, using a downsampling rate of 0.5 produced a set of 1,024 points.
%
Comparing the orange plot with the gray and blue plots in Figure~\ref{fig:distinctive_test}, we can see that having more points on the distinctive regions can better preserve the discrimination of the shapes, leading to higher shape classification accuracy, particularly for results with fewer points.
%
%Particularly, even we halved the total number of sample points, the overall accuracy drops only very little (around 0.9\%), if we deliberately preserve more points on the higher distinctive regions.
Hence, this quantitative comparison shows that the distinctive regions detected by our method are helpful to the classification of 3D shapes.
\NEW{Additionally, since the orange plot is above the gray plot, this indicates that not all high-curvature regions are important for recognition and our method does not simply take all regions of specific curvature profiles,~\eg, sharp corners and extremities, as distinctive.}

\subsection{Effect of using Different Training sets}
\label{subsec:database}

A notable characteristic of detecting distinctive regions is that the results highly relate to the given set of shapes, or the {\em training set\/}.
To verify this, we conducted an experiment to explore the effect of training set as follows.
First, we collected two training sets:
(i) 500 four-engine airplanes and 250 tail-engine airplanes (see Figure~\ref{fig:database} (top-left) for examples), and
(ii) we kept the four-engine airplanes but replaced the tail-engine airplanes with around 1,000 two-engine airplanes (see Figure~\ref{fig:database} (bottom-left) for examples).
Please see Supplementary Material Part E for more examples in the datasets.
Further, we trained our network on each training set separately with $C$ set to two, and employed the two trained network models to detect distinctive regions on four-engine airplanes.

Figure~\ref{fig:database} (right) shows the distinctive regions detected on four different four-engine airplanes.
The interesting observation is that when trained with the four-engine vs. tail-engine airplane dataset (top row), our network tends to highlight all the four engines on the test airplanes.
On the other hand, when trained with the four-engine vs. two-engine airplane dataset (bottom row), our network tends to only highlight the outer two engines on the airplanes as distinctive regions.
\NEW{\NEWNEW{In particular}, the four test airplanes have different size, engine shape, and wing shape, where the middle two have special structures,~\ie, extra fuel tanks and a radar on top.
Yet, given these shapes as inputs, our method still detects consistent distinctive regions.}
%Furthermore, if we look closer to the top training set, the two kinds of airplanes not only have different engine locations but also different tail shapes; yet, our trained network can weakly highlight the tails on the test airplanes (see Figure~\ref{fig:database} (top-right)), even these regions are not as dominant as the engines.
\NEW{Further, if we look closer to the training set on top left, the two kinds of airplanes not only have different engine locations but also different tail shapes. Even though these tail regions are not as dominant as the engines, our trained network can still weakly highlight the tails on the test airplanes; compare the tails of the test airplanes on top-right vs. bottom-right in Figure~\ref{fig:database}.}
%\lqyu{Maybe we can add more examples of the selected airplane datasets in the supp. to show diversity.}
%
\NEW{Please see Supplementary Material Part L for results on other datasets.}

%%%%%%%%%%%%%%%%%%%%%%%%%%%%%%%%
\begin{figure*}[t]
	\centering
	\includegraphics[width=0.99\linewidth]{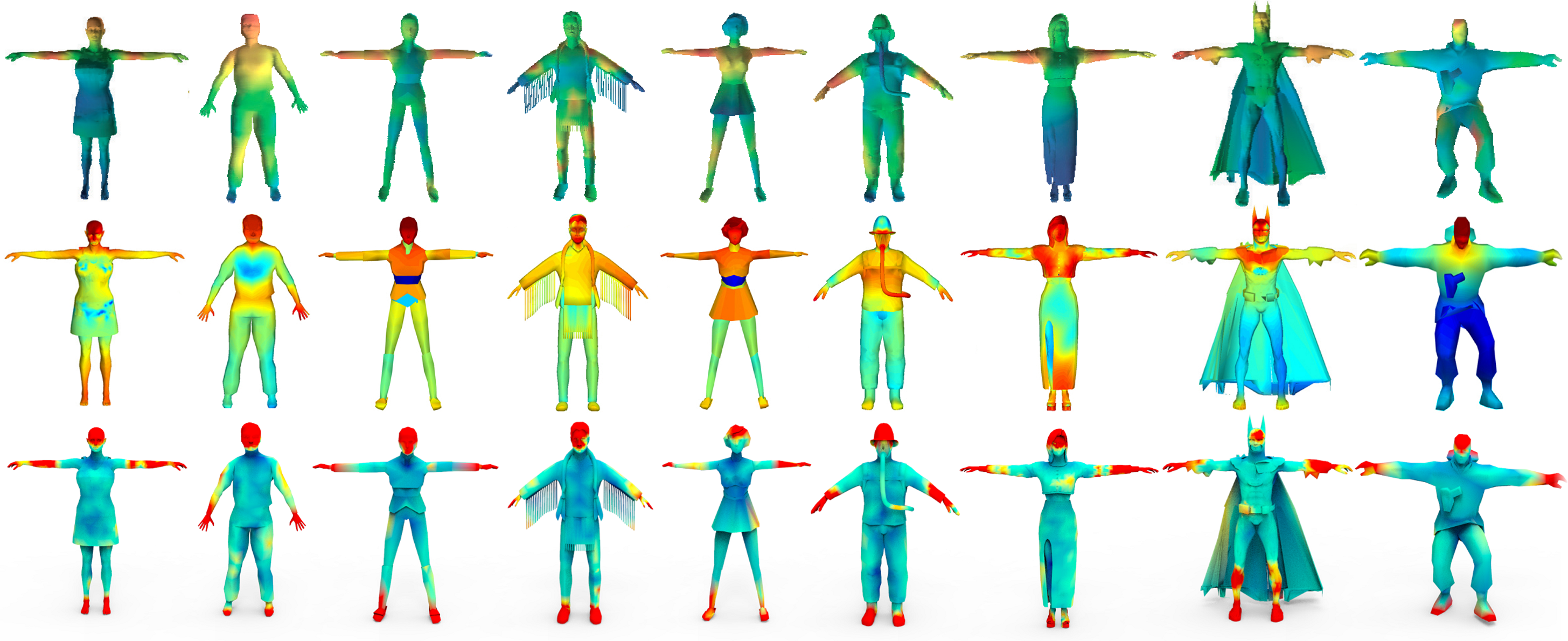}
	\vspace*{-3mm}
	\caption{Distinctive regions detected by Shilane et al.~\shortcite{shilane2007distinctive} (top row), by Song et al.~\shortcite{song2018distinction} (middle row), and by our method (bottom row).
	%\NEW{We train our network using the same training set (\ie, the Princeton Shape Benchmark~\cite{shilane2004princeton}) as the other two, and test the performance of each method on the same testing set.
	%The user study reported in Section~\ref{subsec:comparison} shows that our distinctive regions better match the user preference than the other two.}
	\NEWNEW{For all the three methods, distinctive regions on the person shapes are detected by comparing the person shapes with shapes of other different class types.}
	}
	%the are the best compared with others based on the participants' personal perception.}}
	%\caption{Comparison of distinctive regions detected by different methods. Top: Shilane~\etal's method~\shortcite{shilane2007distinctive} based on hand-crafted features. Middle: Song~\etal's method~\shortcite{song2018distinction} based on supervised MRF-CNN. Bottom: our method.}
	\label{fig:comparison_distinction}
	%\vspace*{-1mm}
\end{figure*}

\subsection{Comparing with Other Methods}
\label{subsec:comparison}

Figure~\ref{fig:comparison_distinction} shows the distinctive regions detected by a traditional method~\cite{shilane2007distinctive} (without deep neural networks), by a weakly-supervised deep learning method~\cite{song2018distinction}, and by our method.
\NEWNEW{We train our network using the same training set,~\ie, the Princeton Shape Benchmark~\cite{shilane2004princeton} as in ~\cite{shilane2007distinctive}, and most class types of training samples in~\cite{song2018distinction} are the same as ours.
For all the three methods, distinctive regions on the person shapes are detected by comparing the person shapes with shapes of other different class types in the data.}
The results of~\cite{shilane2007distinctive} and~\cite{song2018distinction} were directly acquired from their papers.
%, whereas our results were produced using the network trained 
%for
%\phil{on} 
%the ModelNet40 dataset~\cite{wu20153d}.
%
%Moreover, all Person shapes shown here are acquired from the Princeton Shape Benchmark~\cite{shilane2004princeton}.

From the results, we can see that~\cite{shilane2007distinctive} tends to highlight elbows as distinctive regions and ignore semantic parts such as heads and feet, due to the limited representation capability of the hand-crafted features.
For~\cite{song2018distinction}, it is able to highlight heads and hands as distinctive.
Compared with~\cite{song2018distinction}, even though our method is unsupervised, it can detect not only the heads and hands but also the feet as distinctive.
\NEW{\NEWNEW{In particular}, our method detects consistent distinctive regions on these Person shapes, even they have different poses and shapes.}

\subsection{Unsupervised vs. Weakly-supervised Learning}
\label{sub:supervise}
\begin{figure}[!t]
	\centering
	\vspace*{-1mm}
	\includegraphics[width=0.96\linewidth]{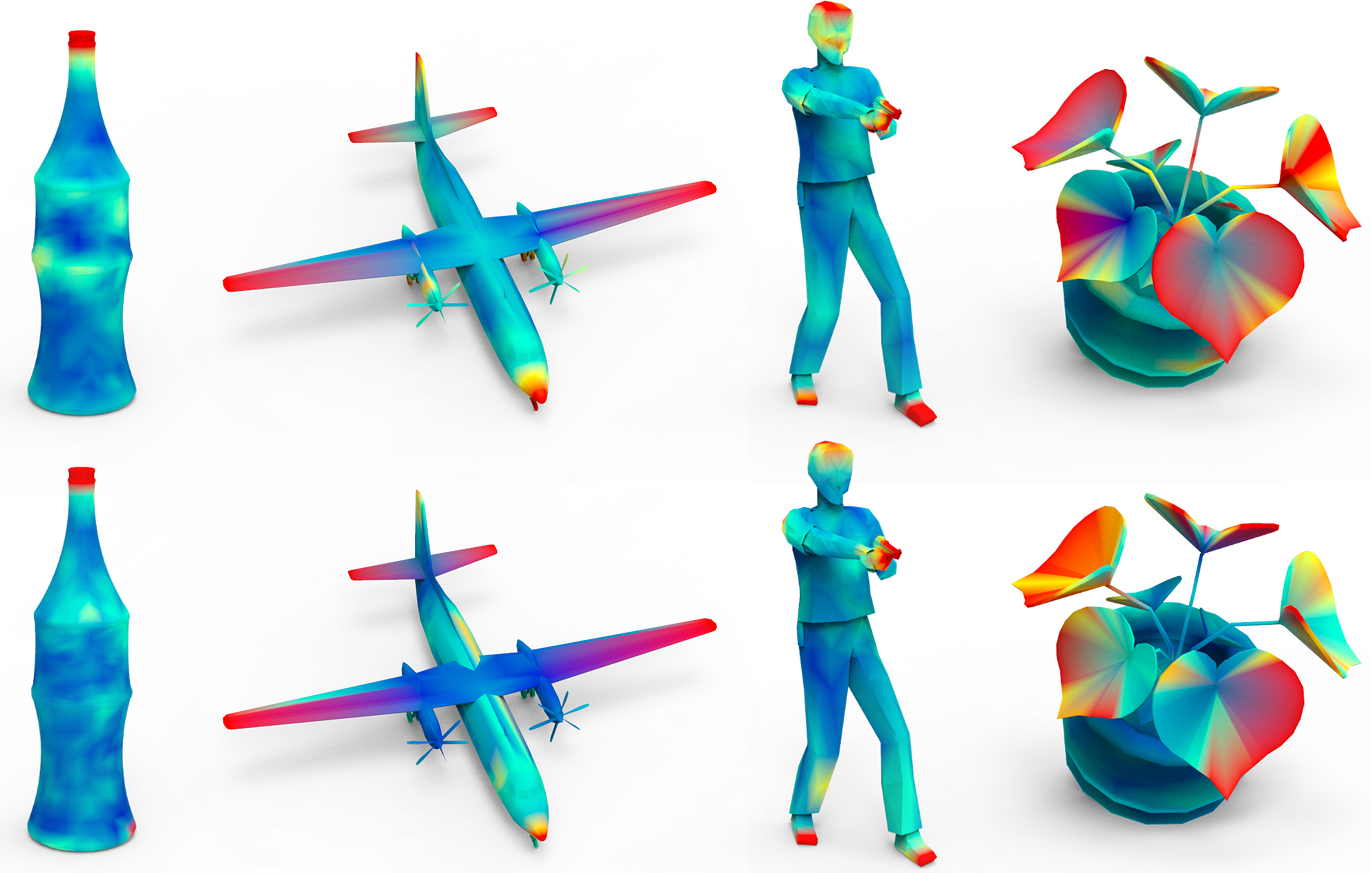}
	\vspace*{-2.5mm}
	\caption{Distinctive regions detected by our method (bottom row) and a weakly-supervised version of our method (top row).
	\NEW{Even without using any class label, our method, which is unsupervised, can still detect regions that are similar to those by a weakly-supervised version of our method.}}
	\label{fig:supervise}
	%\vspace*{-2mm}
\end{figure}

To explore if our unsupervised network can meaningfully cluster the shapes for detecting distinctive regions, we further compared it with a weakly-supervised version of our method.
Specifically, we used the class labels provided in ModelNet40, added a fully-connected layer with 40 output neurons after the global feature (see Figure~\ref{fig:framework}) to regress the class scores, then used the cross entropy loss to replace the unsupervised loss to train this weakly-supervised network.
%
%Figure~\ref{fig:supervise} shows the distinctive regions detected by the supervised (top) and unsupervised (bottom) versions of our method.
From the results presented in Figure~\ref{fig:supervise}, we can see that most distinctive regions detected by the weakly-supervised network (top) can also be found by the unsupervised network (bottom); the unsupervised network only misses a few of them,~\eg, some leaves in Plant.
%our unsupervised network, without using the class labels, can find distinctive regions similar with those from the supervised network, although it misses some regions,~\eg, the left hand of the person and the armrest of the chair.
This comparison result gives evidence that even without the class labels, the performance of our unsupervised method is still comparable to that of the weakly-supervised version of our method.

%%%%%%%%%%%%%%%%%%%%%%%%%%%%%%%%

Further, we quantitatively compare the two results by following~\cite{dutagaci2012evaluation} to compute the False Negative Error (FNE) and False Positive Error (FPE).
Specifically, given distinctiveness values $d_{i,j}$ and $\widehat{d}_{i,j}$ at point $\mathbf{p}_{i,j}$ detected by the unsupervised and weakly-supervised networks, respectively, we first located two sets of more distinctive points per shape $S_j$ by a threshold $d_t$:
$         Q_j  = \{\mathbf{p}_{i,j} |           d_{i,j}>d_t\}$ and
$\widehat{Q_j} = \{\mathbf{p}_{i,j} | \widehat{d}_{i,j}>d_t\}$.
%where $d_t$ is a threshold value set to be \phil{0.2} \xz{actually, 0.2 is not a very low value, since most of the points have the distinctiveness value lower than 0.1. Here, I don't want to write the detailed value, because we have to explain a lot, otherwise, reviewers may think that we set a too low threshold}.
%
By regarding $\widehat{Q}_j$ as the {\em ground truth\/}, a point $\widehat{\mathbf{q}} \in \widehat{Q}_j$ is said to be {\em covered\/} by $Q_j$, if there exists point $\mathbf{q} \in Q_j$, such that $||\widehat{\mathbf{q}}-\mathbf{q}||_2 \leq rD_j$ and $\mathbf{q}$ is not closer to any other point in $\widehat{Q}_j$,
where $D_j$ is the bounding sphere diameter of shape $S_j$ and $r$ is a parameter (ratio) to control the localization tolerance.
\NEW{Then, we compute
$\text{FNE}_j=(|\widehat{Q}_j|-N_c)/|\widehat{Q}_j|=1-N_c/|\widehat{Q}_j|$, where $N_c$ is the number of points in $\widehat{Q}_j$ that are covered by $Q_j$.
%, so $|\widehat{Q}_j|-N_c$ is the number of points in $\widehat{Q}_j$ that are not covered by any point in $Q_j$.
On the other hand, each covered point in $\widehat{Q}_j$ corresponds to a unique point in $Q_j$, so the points in $Q_j$ that are without any correspondence in $\widehat{Q}_j$ are regarded as false positives.
Hence, $\text{FPE}_j=(|Q_j|-N_c)/|Q_j| = 1 - N_c/|Q_j|$.}
% to measure the deviation between $Q_j$ and $\widehat{Q}_j$, 
%where 
%$N_c$ is the number of points in $\widehat{Q}_j$ that are covered by $Q_j$.
Here, $\text{FNE} \in [0,1]$, $\text{FPE} \in [0,1]$, and a small value indicates high consistency between $Q_j$ and $\widehat{Q}_j$.

\begin{figure}[!t]
	\centering
	\includegraphics[width=0.99\linewidth]{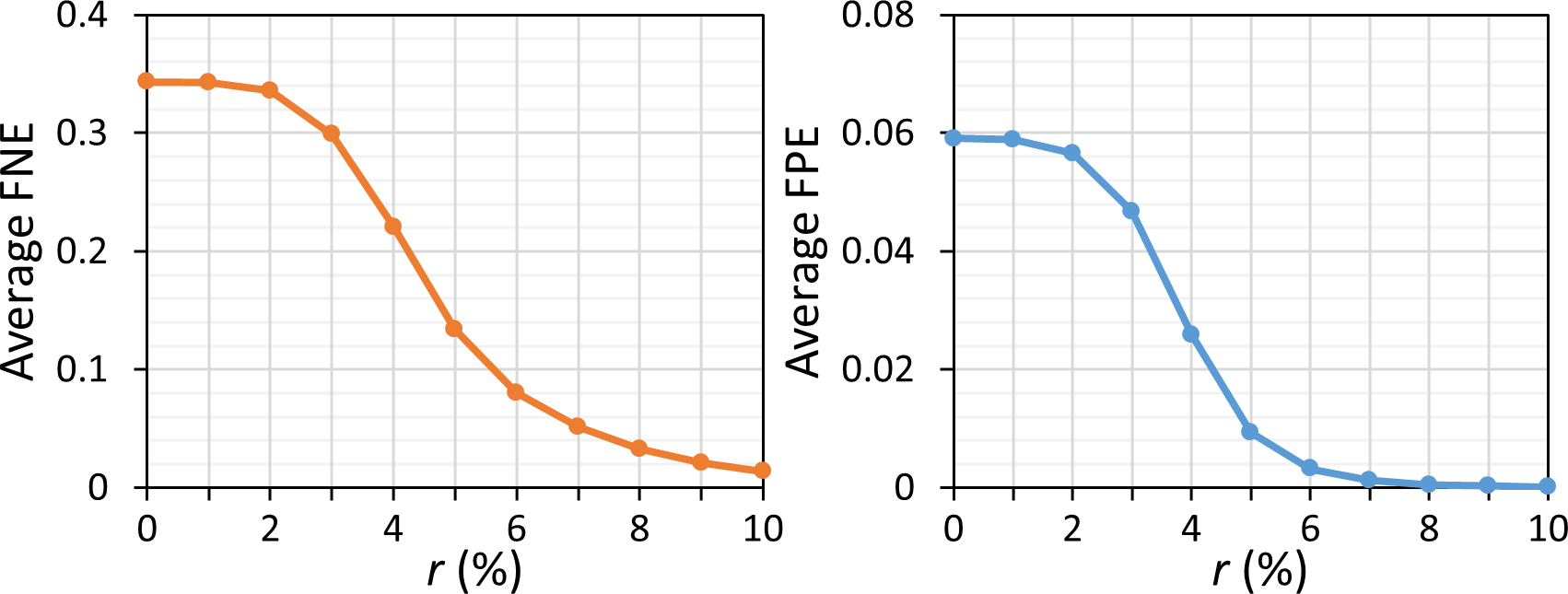}
	\vspace*{-2.5mm}
	\caption{Average false negative error (FNE) and false positive error (FPE) over 400 models plotted against tolerance parameter $r$.}
	%, which is the percentage relative to the bounding sphere diameter.}
	\label{fig:supervise_quantify}
	%\vspace*{-1mm}
\end{figure}

Figure~\ref{fig:supervise_quantify} plots the FNE and FPE values averaged over 400 randomly-selected objects (from the 40 different classes in the ModelNet40 testing split) against $r$.
We can see that when $r$ is just around 8\% to 10\%, both FNE and FPE are very close to zero, thus demonstrating the {\em high consistency between the distinctive points\/} detected by the two networks.
Particularly, the average FPE is very low even when $r=0$, meaning that most of the distinctive regions detected by our unsupervised method are also the distinctive regions detected by the weakly-supervised version of our method.

%%%%%%%%%%%%%%%%%%%%%%%%%%%%%%%%

\begin{figure}[!t]
	\centering
	\includegraphics[width=0.99\linewidth]{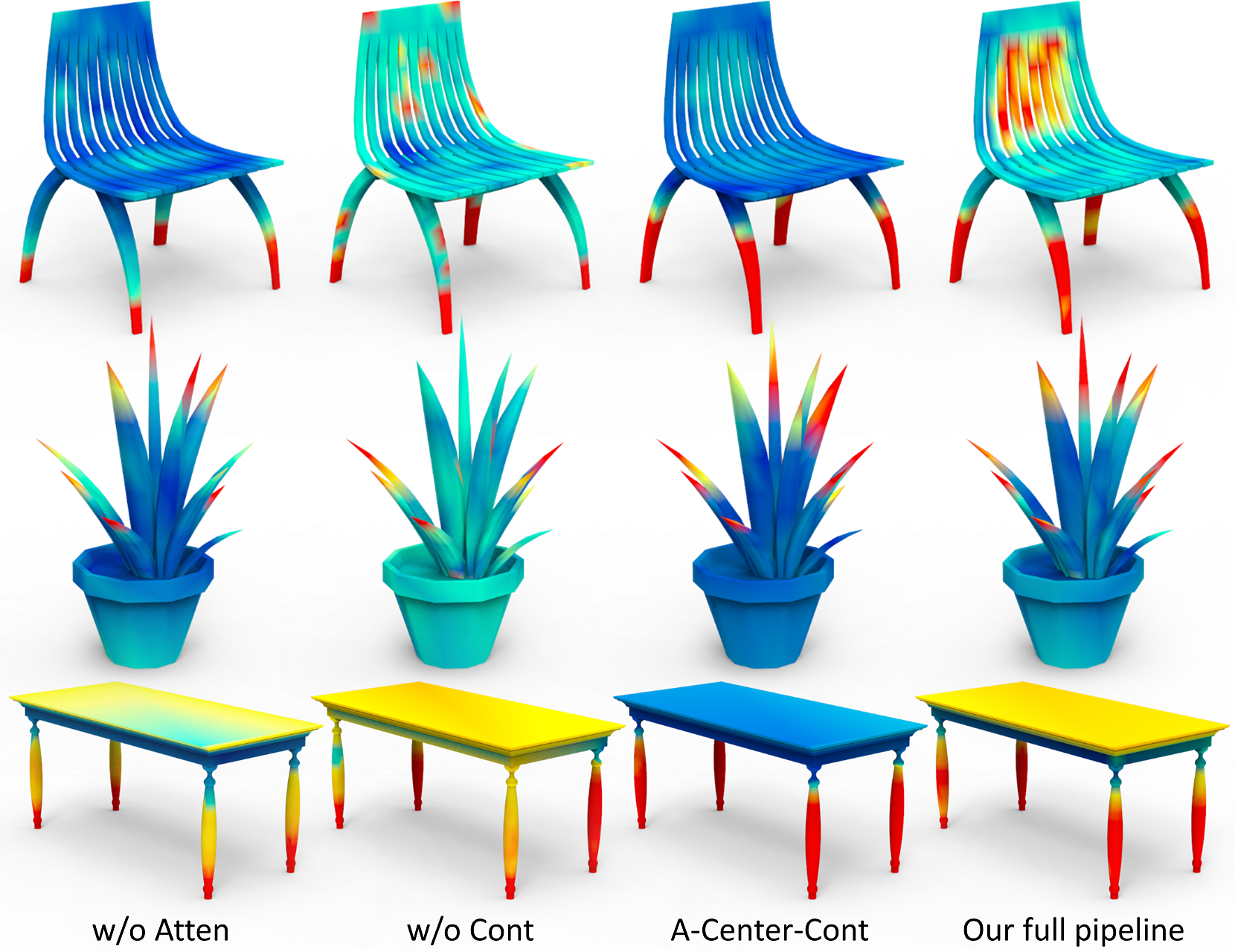}
	\vspace*{-2mm}
	\caption{Visual results in ablation study.}
	\label{fig:comparison}
	\vspace*{-2mm}
\end{figure}

\subsection{Ablation Study}

Next, we analyzed the major elements in our network by removing or replacing each of them when we train our network on the ModelNet40 models:
(i) \textit{w/o Atten} -- removing the channel-spatial attention unit (see Figure~\ref{fig:attention} for details);
(ii) \textit{w/o Cont} -- removing the $L_{contrastive}$ term from the joint loss in Eq.~\eqref{equ:totalloss}; and
(iii) \textit{A-Center-Cont} -- replacing $L_{cluster}$ in Eq.~\eqref{equ:cluster_softmax} with an adapted center loss~\cite{wen2016discriminative} for unsupervised clustering-based learning: $\frac{1}{2}\sum_{j=1}^{N_\text{obj}}$$\parallel$$\mathbf{g}_j-\bar{\mathbf{g}}_{q}$$\parallel^2$ with a goal of minimizing the intra-cluster variations, while keeping the features of different clusters separable.

Figure~\ref{fig:comparison} shows the distinctive regions detected by our method under the four different settings.
Comparing the results produced with the \textit{w/o Atten} setting (left-most) and with our full pipeline (right-most), we can see that the attention unit helps locate distinctive regions that span over a larger spatial areas,~\eg, the Chair's back.
Looking at the results produced with \textit{w/o Cont} and \textit{A-Center-Cont}, we can see that they tend to miss some important regions,~\eg, the Table's desktop, or contain noise,~\eg, the Chair's back.

\newcommand{\BE}[1]{{\textbf{#1}}}
\begin{table}
	%	\caption{Comparison of the overall shape classification accuracy on ModelNet40 under increasing downsampling rate. We %down-sample testing point sets by dropping more distinctive points with the distinctiveness produced by different  methods. The %value in parentheses indicates the percentage of accuracy reduction compared with the classification accuracy without dropping.}
	%
	\caption{Comparing the overall shape classification accuracy on ModelNet40 when downsampling the test shapes with more points on the distinctive regions detected under different settings.}
	\label{tab:comparison}
	\vspace*{-3mm}
	\centering
	\begin{center}
		\scalebox{0.85}
		{
			\begin{tabular}{c|c|c|c|c|c|c} \toprule[1pt]
				Different	& \multicolumn{6}{c}{Number of downsampled points} \\
				\cline{2-7} 
				settings	&1024	       &512    	      &256	         &128	         &64	         &32		\\
				\hline
				\hline
				w/o Atten
				&0.877		&0.871		&0.806		&0.648		&0.402		&0.244		\\
				\hline
				w/o Cont
				&0.878	       &0.871		&0.808		&0.653		&0.413		&0.245		\\
				\hline
				A-Center-Cont
				&0.876		&0.865		&0.818		&0.658		&0.405		&0.235		\\
				\hline
				Our full pipeline
				&\BE{0.881}	&\BE{0.874}	&\BE{0.829}	&\BE{0.729}	&\BE{0.550}	&\BE{0.408}	\\
				\bottomrule[1pt]
			\end{tabular}
		}
	\end{center}
	\vspace*{-1mm}
\end{table}

Besides visual comparison, we performed the same quantitative evaluation on the results here, as in Section~\ref{subsec:distinctive}.
Table~\ref{tab:comparison} shows the shape classification accuracy on the ModelNet40 test dataset for the four different settings in terms of decreasing number of downsampled points.
From the table, we can observe that our full pipeline leads to the highest classification accuracy.
Since we preserve more points on distinctive regions, this means that the three network elements being explored in this experiment all contribute to improve the detection of the distinctive regions.

\subsection{User Studies}
\label{subsec:user_study}

To obtain a sense of how consistent our results are with humans, we conducted two user studies, which we shall elaborate below.
%To explore how our network predictions are consistent with human perception, we conducted two kinds of user studies, which we shall elaborate below.
%
\NEW{The key idea behind the studies is that we try to simulate the network clustering process with humans to obtain the distinctive regions of some test shapes.
Specifically, we started by introducing the definition of distinction to each participant to confirm that all participants understood the meaning of distinction.
Then, we showed the training dataset to the participants.
However, to avoid fatigue, we randomly selected a subset of 3D shapes from different classes in the training dataset, and showed these shapes to each participant on a computer display, on which the participant can rotate each shape and explore its details.
Please refer to Supplementary Material Part F for some screenshots.
After that, each participant was given a set of test shapes, and asked to cluster the shapes and label the distinctive regions on the shapes that affect how they cluster.
All the shapes are presented simply in a colorless manner to the participants in both of the studies.}
%were given a set of 3D shapes shown together on a computer display, where they can rotate individual shapes; see Supplementary Material part F for the interface screenshots.
%They were then asked to cluster the shapes and label the distinctive regions on shapes that affect how they cluster.

%%%%%%%%%%%%%%%%%%%%%%%%%%%%%%%%

%\vspace*{-5pt}
\paragraph{Intra-class prediction.}
The first user study explores how humans find distinctive regions on shapes of the same class.
Here, we employed 
(i) the dataset of four-engine vs. two-engine airplanes; and 
(ii) the dataset of four-engine vs. tail-engine airplanes,
as presented in Section~\ref{subsec:database}; see again Figure~\ref{fig:database} (left).
To avoid bias due to the dataset similarity, we randomly divided the participants into two groups, one for each set.
For the first group, we randomly selected 10 four-engine airplanes and 22 two-engine airplanes, and presented them in random order on a computer display.
Then, the participants were asked to divide the 32 airplanes into two clusters and label the distinctive regions on the four-engine airplanes.
For the other group of participants, we randomly selected 10 four-engine and 22 tail-engine airplanes from the other dataset, and performed the same procedure with the participants.
%by randomly selecting 22 two-engine airplanes and 10 four-engine airplanes.
%
%As we carried out this user study, we found it fairly easy for humans to detect intra-class distinctive regions, so we recruited only ten participants in this study.

All ten participants (both groups) recruited in this study clustered the airplanes in the same way as our network.
For the four-engine vs. two-engine dataset, all participants marked the outer two engines as distinctive in the four-engine airplanes.
Their results are the same as our network predictions; see Figure~\ref{fig:database} (bottom-right).
For the four-engine vs. tail-engine dataset, all participants marked the four engines as distinctive in the four-engine airplanes, and their results are almost the same as our network (see Figure~\ref{fig:database} (top-right)), except for the tails of the airplanes; since the tail-engine airplanes mostly have T-shaped tails, which are generally absent in four-engine airplanes.
Without our reminder, only one participant noticed the T-shaped tails, but when we asked the other participants whether such tails are also distinctive, they all strongly agreed.
This study shows that our network is able to attend to large and small distinctive regions, which may even be overlooked by humans.

%%%%%%%%%%%%%%%%%%%%%%%%%%%%%%%%

\begin{figure}[!t]
	\centering
	\includegraphics[width=0.99\linewidth]{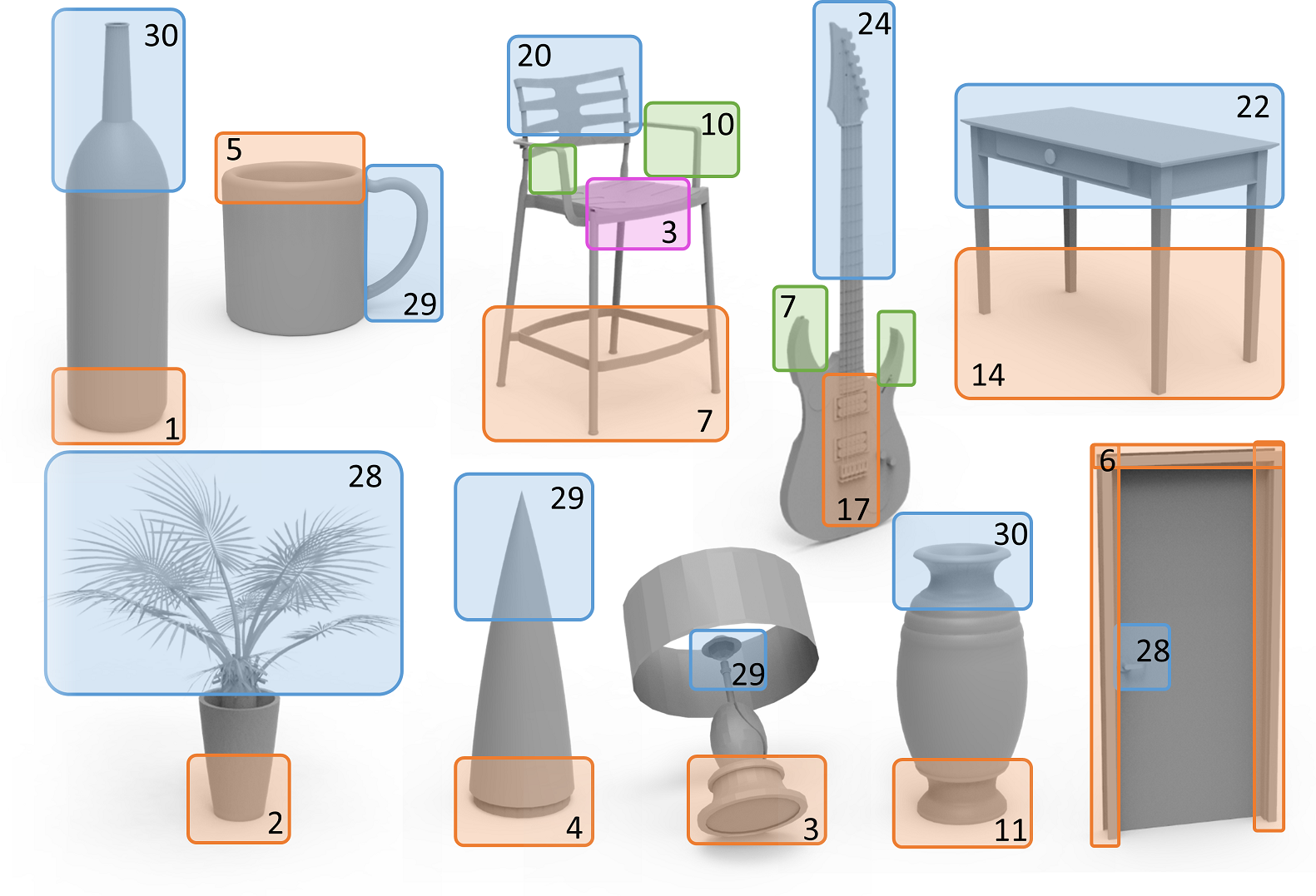}
	\vspace*{-2mm}
	\caption{The ten objects on which the participants mark distinctive regions.
		The above boxes reveal the participant-marked distinctive regions and the corresponding number of participants marked on each region.
		Note that each participant may mark more than one region on the same object.}
	%	\caption{The eight objects that users need to label distinctive regions and their corresponding key points detected by our method.}
	\label{fig:keypoint}
	\vspace*{-1mm}
\end{figure}

%for the two-engine vs. four-engine dataset, all participants cluster the 32 airplanes in the same way as in our dataset and they all pointed out the outer two engines as the unique feature in the four-engine airplanes. Their results are the same as our network predictions (see Figure~\ref{fig:database} (top-right)).}
%
%For the tail-engine vs. four-engine dataset, \xz{all participants still cluster these airplanes successfully} and they all pointed out the four engines as the unique feature in the four-engine airplanes.
%Their results here are almost the same as our network (see Figure~\ref{fig:database} (bottom-right)), except for the tail parts in the airplanes, since airplanes with tail engines \phil{mostly} have T-shaped tails.
%Without our reminder, the participants not notice the tail part, but when we asked them whether the tail is also distinctive, they strongly agreed.
%
%This user study shows our network can detect tiny distinctive areas such as the airplane tails, which may be ignored by humans.
%This user study shows that our network prediction trained with an intra-class database is highly consistent with human perception.
%Particularly, our network can detect some tiny distinctive areas (\eg, the tail region) that may be ignored by human beings at %the first glance given many objects. 

%%%%%%%%%%%%%%%%%%%%%%%%%%%%%%%%

\begin{table}[!t]
\caption{Quantitative evaluation on the inter-class prediction consistency between our method and the participants.}
%All three metrics ranges in $[0,1]$, where a lower value indicates a higher consistency between the predictions.}
\label{tab:user_study}
\vspace*{-3mm}
\begin{center}
\scalebox{0.78}
	{
\begin{tabular}{@{\hspace{1mm}}c@{\hspace{1mm}}|c@{\hspace{2mm}}c@{\hspace{2mm}}c@{\hspace{2mm}}c@{\hspace{2mm}}c@{\hspace{2mm}}c@{\hspace{2mm}}c@{\hspace{2mm}}c@{\hspace{2mm}}c@{\hspace{2mm}}c@{\hspace{2mm}}|@{\hspace{1mm}}c}
\toprule[1pt]
	& Bottle & Chair & Cone  & Cup & Door  & Guitar & Lamp & Plant & Table & Vase & Avg \\ \hline
FNE	& 0.02        & 0.40   & 0.08 & 0.10 & 0 & 0.51 & 0.07   & 0.07      & 0 & 0 & 0.13	\\
FPE	& 0     & 0.53   & 0.03 & 0.03    & 0.40    & 0.20    & 0.03   & 0.07	  & 0.35 & 0.30 & 0.19	\\
WME	& 0.03        & 0.33   & 0.12 & 0.15 & 0.18 & 0.50 & 0.09   & 0.06      & 0 & 0 & 0.15	\\
\bottomrule[1pt]
\end{tabular}
}
\end{center}
\vspace*{-1mm}
\end{table}

\begin{figure*}[t]
	\centering
	\includegraphics[width=0.99\linewidth]{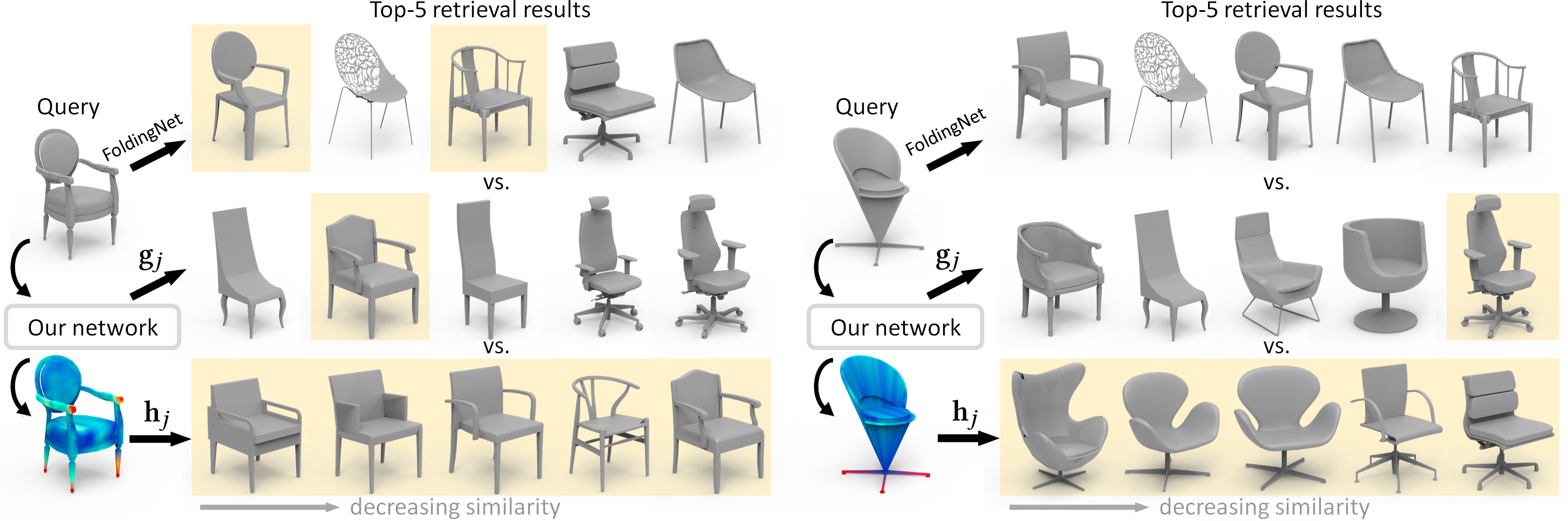}
	\vspace*{-2mm}
	\caption{Two sets of shape retrieval results (left \& right).
	In each set, we show the top-five similar shapes retrieved by using the per-shape global feature from FoldingNet~\cite{yang2018foldingnet} (top) and from our network,~\ie, $\mathbf{g}_j$ (middle), and by using our distinctiveness-guided global feature $\mathbf{h}_j$ (bottom).
		Retrieved shapes of substructures (armrest (left) \& swivel base (right)) similar to the query shape are marked over a yellow background.}
	%\phil{maybe do an inverse, highlight the retrieved shapes that are more similar by a yellow background instead}}
	%
	%\caption{Top-five shape retrieval results produced using GASD (top), our general retrieval method using $\mathbf{g}_j$ (middle), %and our distinctiveness-guided retrieval method using $\mathbf{h}_j$ (bottom).
	%Retrieved shapes of substructures very different from the query shape are marked over a yellow background.}
	%\caption{Top-5 retrieval results given two kinds of chairs as queries by employing GASD method (the top row), our general retrieval method (the middle row) and our distinctiveness-guided retrieval method (the bottom row). Incorrect retrieval shapes are marked with yellow background.}
	\label{fig:retrieval}
	\vspace*{-1.5mm}
\end{figure*}

%%%%%%%%%%%%%%%%%%%%%%%%%%%%%%%%

%\vspace*{-5pt}
\paragraph{Inter-class prediction.}
The second study explores how humans find distinctive regions between shapes of various kinds.
Here, we employed the models from ModelNet40 and recruited 30 participants.
To avoid fatigue, we randomly selected 75 shapes evenly from 15 different classes, and further selected ten objects of different classes from the set; see Figure~\ref{fig:keypoint}.
Then, we showed the 75 shapes to each participant and asked him/her to mark distinctive regions on the ten selected objects.
Particularly, we explained the definition of distinctiveness,~\ie, the distinctive regions should be common in each specific class, while being unique {\em relative\/} to other classes.
%of models in Figure~\ref{fig:vis}.
%\phil{say it step by step... talk about first the dataset in use? just say Fig.5 is not clear enough}

During the study, we found that the size of the marked region varies among the participants, even at the same object location.
Also, they focus more on the shape features than on the size of the marked regions.
%on the objects.
%
Figure~\ref{fig:keypoint} shows the summary of human marked regions on the ten objects.
In each marked region, the corresponding number indicates how many participants marked on the same area. 

To quantitatively compare the distinctive regions marked by the participants and detected by our method, we adapted the FNE and FPE metrics (see Section~\ref{sub:supervise}) as follows.
%
%First, we followed the outward statistical testing method in~\cite{wang2016automatic,shu2018detecting} to automatically cluster the distinctive regions detected by our network on each of the eight objects into {\em extremities\/} that represent the distinctive regions; see Figure~\ref{fig:keypoint}.
First, for regions detected by our network, we set a threshold to keep only the high-distinctive regions as the final detected distinctive regions; see Figure~\ref{fig:vis} for the distinctiveness distribution on the ten objects.
Intuitively, we keep only the yellow and red regions on the objects.
Next, we compared these regions with the regions marked by each participant, and regard his/her marked regions as the ground truth, where a marked region is said to be {\em covered\/}, if both the participant-marked region and the network-detected region cover almost the same structures.
On the other hand, the network-detected regions that are covered by the participant-marked regions are said to be the true positives.
%it contains an extremity detected by our network.
%On the other hand, extremities that fall inside the marked regions are said to be true positives.
%
\NEW{Hence, for each ($k$-th) participant, we define
$N^c_{j,k}$ as the number of participant-marked regions that are covered and 
$T^h_{j,k}$ as the total number of participant-marked regions, on the $j$-th object.
Also, we define
$T^d_j$ as the total number of network-detected regions on the $j$-th object.
Then, similar to Section~\ref{sub:supervise}, we compute the corresponding 
$\text{FNE}_{j,k} = 1 - N^c_{j,k} / T^h_{j,k}$ and 
$\text{FPE}_{j,k} = 1 - N^c_{j,k} / T^d_j$, and 
further compute the FNE and FPE values averaged over all the participants per object.}
Additionally, to account for the frequently-marked regions, we adopted the Weighted Miss Error (WME) metric~\cite{dutagaci2012evaluation},~\ie,
$\text{WME} = 1 - \sum_k N^c_{j,k}/\sum_k T^h_{j,k}$.

Table~\ref{tab:user_study} shows the per-object FNE, FPE, and WME values, as well as their overall averages over the ten objects.
All three metrics range $[0,1]$, where a low value indicates high consistency between the participant-marked and network-detected regions.
From the table, we can see that most values are very low and several are even zeros, indicating that most participant-marked distinctive regions can be detected by the network, and vice versa.
However, values for some complex objects are a bit higher,~\eg, the Chair and the Guitar, since different participants may mark on different structures.
%We observed that for Car, most participants did not rotate and mark the back side, while for Person, most participants only mark one or two most obvious regions,~\eg, the head of the Person shape.
Yet, the overall values are very close to zeros, meaning that the distinctive regions detected by our network are highly consistent with the distinctive regions marked by the participants.
%\phil{Pls check the participant data and revise later}
\NEW{Furthermore, we also conducted a statistical test to show that 30 participants are sufficient to produce stable human labels in the analysis.
Please refer to Supplementary Material Part F for the details.}

%% file: applications.tex
\section{Applications}
\label{sec:applications}

\NEW{Being able to discover distinctive regions on 3D shapes enables us to support and enhance various applications, such as
shape retrieval~\cite{shilane2006selecting,gal2006salient},
%\phil{XZ: to add}
shape simplification~\cite{garland1997surface},
remeshing~\cite{alliez2002interactive},
best view selection~\cite{shilane2007distinctive},
perception-aware 3D printing~\cite{zhang2015perceptual}.
In this section, we present three typical applications to demonstrate the applicability of our technique in distinctiveness-guided shape retrieval, sampling, and view selection in 3D scenes.}

%Discovering the distinctive regions on a 3D shape with respect to the unlabeled database enables a variety of applications. 
%In this section, we present three prototype applications demonstrating the applicability of our technique in shape retrieval, distinctiveness-driven re-sampling, and distinctive region detection in 3D scenes. 

%%%%%%%%%%%%%%%%%%%%%%%%%%%%%%%%%%%%%%%%%%%%%%%%%%%%%%%%%%%%

\subsection{Distinctiveness-guided Shape Retrieval}
\label{subsec:retrieval}

Conventional approaches for shape retrieval first extract representative shape descriptors then retrieve similar shapes based on the distance between the extracted descriptors; see~\cite{tangelder2004survey} for a survey.
Though these approaches have achieved promising performance for inter-class retrieval, they tend to have limited ability for fine-grained intra-class retrieval, since the extracted descriptors are global.
With our detected distinctive regions, we can perform {\em distinctiveness-guided shape retrieval\/}, which enables fine-grained intra-class shape retrieval,~\eg, retrieving swivel chairs from a large collection of chairs; see Figure~\ref{fig:retrieval} (right).

The main idea is to only consider these distinctive point features rather than regarding all the point features equally.
%The general strategy is to take the network (per-shape) features as shape descriptors for the retrieval.
%The general strategy is to represent a shape by its feature with the guidance of the distinctiveness values learnt from the network.
%
%In detail, we randomly sample $N$ (\eg, 2,048) points on each shape $S_j$ and extract the Fast Point Feature Histograms (FPFH) descriptor~\cite{RusuDoctoralDissertation} feature $\mathbf{h}_{i,j}$ for each point $\mathbf{p}_{i,j}$. 
In detail, we randomly sample $N$ (\eg, 2,048) points on each shape $S_j$ and feed these points into our trained network to obtain local per-point features $\mathbf{F}_j^r$ and global per-shape features $\mathbf{g}_j$; see Section~\ref{subsec:feature_embedding}.
As described in Section~\ref{sec:point_vis}, we further obtain per-point distinctiveness $d_{i,j}$ from $\mathbf{f}^r_{i,j}$.
%by taking the maximum value in $\mathbf{f}^r_{i,j}$.
To facilitate fine-grained intra-class retrieval, instead of directly using $\mathbf{g}_j$ as the representative descriptor, we select only the more distinctive per-point features and average over them to obtain the distinctiveness-guided global feature $\mathbf{h}_j$:
\begin{equation}
\label{equ:distinctive_feature}
\mathbf{h}_j = {\frac{\sum_{i=1}^N \mathbb{I} \{d_{i,j}>\Delta_d\} \mathbf{f}^r_{i,j}}{\sum_{i=1}^N \mathbb{I}\{d_{i,j}>\Delta_d\}}},
\end{equation}
where $\mathbb{I}$ is the indicator function and $\Delta_d$ is the threshold.
Lastly, we measure the similarity between shapes by computing the Euclidean distance between $\mathbf{h}_j$ of the shapes. % (or $\mathbf{g}_j$, in the case without using the distinctiveness values) of the shapes.
%\phil{why use FPFH, a work in 2009? if we use some more powerful descriptors, can they do a better job than ours or at least similar to ours?}
%%%%%%%%%%%%%%%%%%%%%%%%%%%%%%%%%%%%%%%%%%%%%%%%%%%%%%%%%%%%

\begin{figure}[t]
	\centering
	\includegraphics[width=0.99\linewidth]{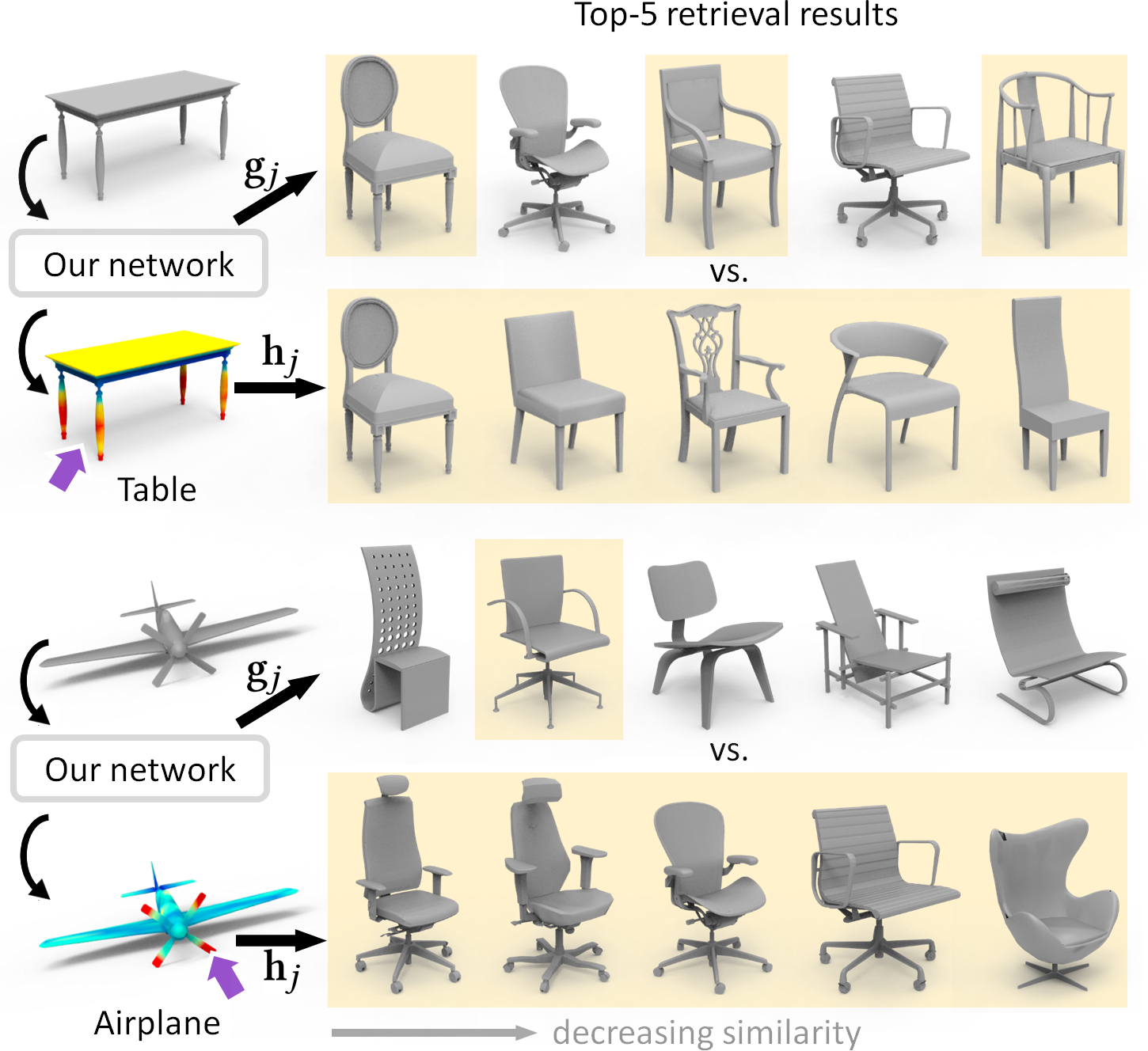}
	\vspace{-2mm}
	\caption{\NEW{Top-five shape retrieval results by using a Table shape with four legs (top) or using an Airplane shape with a propeller (bottom) as the query to search over the Chair dataset.
			Retrieved shapes with similar substructures are marked over a yellow background.}}
	\label{fig:retrieval_spectial}
	\vspace{-2mm}
\end{figure}
%%%%%%%%%%%%%%%%%%%%%%%%%%%%%%%%%%%%%%%%%%%%%%%%%%%%%%%%%%%%

\begin{figure*}[t]
	\centering
	\includegraphics[width=0.99\linewidth]{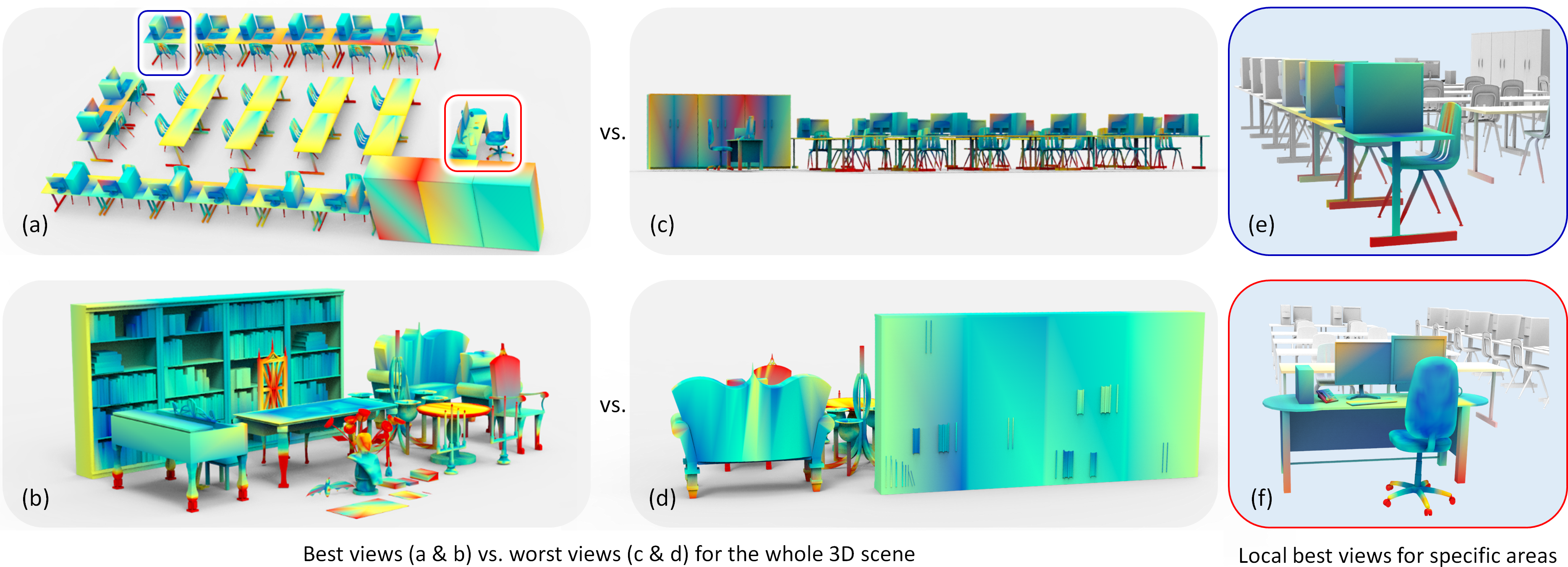}
	\caption{Using the network-predicted distinctiveness values over a scene, we can find best views with maximized distinctiveness.}
	\label{fig:large_scale}
	\vspace{-2mm}
\end{figure*}

\begin{figure}[t]
	\centering
	\includegraphics[width=0.99\linewidth]{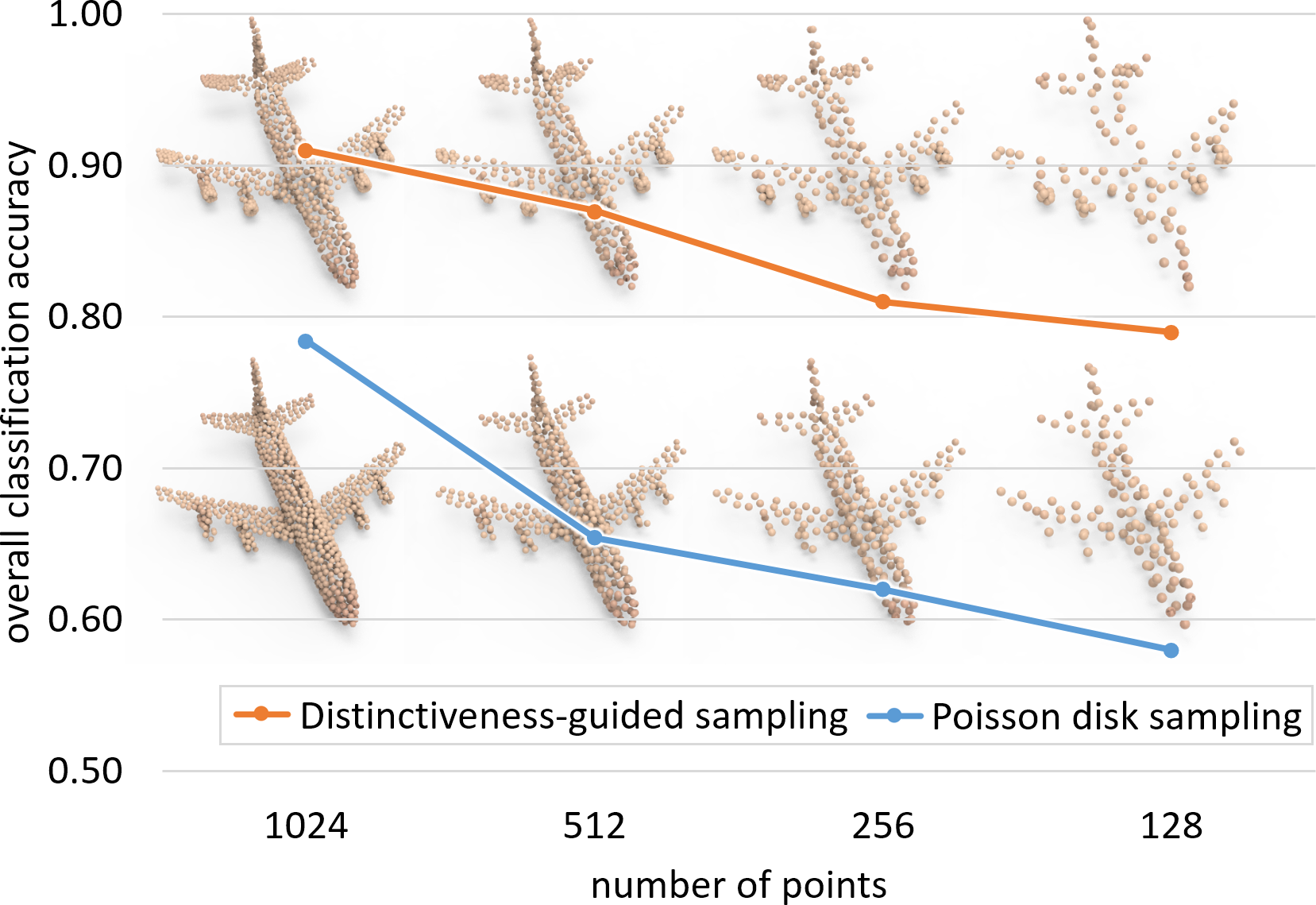}
	\vspace{-2mm}
	\caption{Shape classification accuracy using distinctiveness-guided sampling (top row) vs. conventional Poisson disk sampling (bottom row).}
	%	\caption{Shape classification accuracy on point sets (of various sizes) produced using distinctiveness-guided sampling (top row) vs. conventional Poisson disk sampling (bottom row).}
	%	\caption{Comparison results on different sampling methods in terms of unsupervised overall classification accuracy.\phil{please use some other shapes... too many airplanes in this paper}}
	\label{fig:sampling}
	\vspace{-2mm}
\end{figure}

Figure~\ref{fig:retrieval} shows two sets of results, each using a different chair as the query shape.
As a comparison, we applied a recent unsupervised network FoldingNet~\cite{yang2018foldingnet} to extract per-shape global feature for similarity computation, and the results are shown in the top row.
Besides, we also directly employed our per-shape feature $\mathbf{g}_j$ for retrieval; see the middle row.
In each set, we retrieved the top-five similar shapes from a data pool of 100 different chairs, \NEW{and manually marked in yellow (see the figure) the results with substructures similar to the query shape: 
for the left example in Figure~\ref{fig:retrieval}, we consider a retrieval result as similar to the query shape, if it also has an armrest and four legs as its substructures, while for the right example in Figure~\ref{fig:retrieval}, we consider a retrieval result as similar to the query shape, if it also has a swivel base as its substructures.}
%
%(i) the globally aligned spatial distribution (GASD) descriptor~\cite{do2016efficient},
%(ii) our general shape retrieval using $\mathbf{g}_j$, and 
%(iii) our distinctiveness-guided shape retrieval using $\mathbf{h}_j$.
%%
%Note that for both query shapes, we have carefully examined the database (with 100 chairs) that there are just a few highly-matched shapes (fewer than eight).
%For both query shapes, we have carefully examined the database that among the 100 chairs, there are only seven and nine highly-similar chairs for the left and right query shapes shown in the figure, respectively.
%Hence, it is nontrivial to find exactly these similar chairs.
%
From Figure~\ref{fig:retrieval}, we can see that using the global descriptors, no matter extracted by FoldingNet or our network, the results have very different substructures from the query shapes.
Guided by the network-predicted distinctiveness values, $\mathbf{h}_j$ can help retrieve shapes with more similar substructures (marked in yellow).
\NEW{Please see Supplementary Material Part G for more results.}

\NEW{To further explore the ability of our distinctiveness-guided shape retrieval, we perform shape retrieval on the Chair dataset using a Table and an Airplane as the query.
Figure~\ref{fig:retrieval_spectial} shows the results.
For the top case, the four legs in the Table shape are found to be distinctive, so when using it as the query, four-leg chairs are obtained.
For the bottom case, the Airplane shape has a propeller, so chairs with substructures that look like the propeller (distinctive regions) can be retrieved as the results.
These results demonstrate that our distinctiveness-guided shape retrieval can pay more attention to the local distinctive regions, instead of simply the overall structure.}

\subsection{Distinctiveness-guided Sampling}

\if 0
\begin{table}
	\caption{Comparison on different sub-sampling methods in terms of overall classification accuracy on ModelNet40.}
	\label{tab:down-sample}
	\centering
	\begin{center}
	\scalebox{0.9}
	{
		\begin{tabular}{c|c|c|c|c|c|c} \toprule[1pt]
			\multirow{2}*{method}& \multicolumn{6}{c}{sampling rate} \\
			\cline{2-7} 
			            &0.75	    &0.7    	&0.65	    &0.6 	    &0.55	    &0.5         \\ \hline \hline
			random	    &0.887	    &0.881	    &0.879	    &0.876	    &0.875	    &0.866       \\ \hline
			farthest	&0.880	    &0.880	    &0.878	    &0.878	    &0.877	    &0.874	     \\ \hline
			adaptive	&\BE{0.888} &\BE{0.888} &\BE{0.885} &\BE{0.884} &\BE{0.881} &\BE{0.879}  \\ \bottomrule[1pt]
		\end{tabular}
	}
	\end{center}
\end{table}
\fi

%\phil{distinctive-guided or distinctive-aware here?}

Sampling is a common task in computer graphics, as well as in many domains, for generating point samples to represent a continuous shape.
Using the distinctiveness detected on 3D shapes, we can guide the point sampling process by emphasizing the distinctive regions on the shapes.
In this way, the sampled points can more effectively describe the shapes in terms of discrimination ability.

In our implementation, we take the point distinctiveness values detected by our network to control the local sampling density in an adaptive Poisson disk sampling process.
That is, we set higher sampling density (or equivalent, smaller Poisson disks) for more distinctive regions, and vice versa.
Figure~\ref{fig:sampling} presents the sampling results on a four-engine airplane object with decreasing number of points, where (i) the top row shows the results produced using an adaptive Poisson disk sampling guided by the network-predicted distinctiveness values, and (ii) the bottom row shows the results produced by the conventional Poisson disk sampling, which randomly but uniformly samples the given object.
From the results, we can see that distinctiveness-guided sampling (top) arranges more points in high distinctive regions, such as the engines, thereby enhancing the preservation of the shape's characteristics.
\xz{Please refer to Supplementary Material Part H for more sampling results.}

Furthermore, we compare the two-class classification accuracy for the sampled point sets using our unsupervised network, and present the overall shape classification accuracy as plots in Figure~\ref{fig:sampling}.
In this quantitative comparison, we can further show that the points produced from distinctiveness-guided sampling lead to higher classification accuracy, even with fewer points.

\subsection{View Selection in 3D Scenes}

Given a 3D scene, where are the distinctive regions to attend to?
Using our unsupervised framework, we can find the distinctive regions in an input 3D scene and locate the best views to look at these regions.
Here, we define the best views as those with maximized distinctiveness displayed in the views.

To find such views, we first sample local patches of 2,048 points in the input scene, and feed these patches as inputs to our network to predict per-point distinctiveness.
Since our network is trained on individual objects, we crop regions of around one-meter diameter in the scene for point sampling.
Then, we can combine the results from the local patches to obtain distinctiveness over the scene.
Next, we generate a set of candidate views by uniformly sampling 50 different views on the upper hemisphere that bounds the scene, and evaluate the quality of each view by averaging the distinctiveness over the visible points in the view.
Lastly, we choose the view with the highest averaged distinctiveness as the best view.

Figures~\ref{fig:large_scale} (a) \& (b) show the best views that were automatically selected for two different scenes (courtesy of 3D Warehouse~\shortcite{warehouse}), where the camera was set to look at the center of the scene when sampling the candidate views.
As a comparison, we also show the corresponding worst views for the two scenes; see (c) \& (d).
From these results we can see that, the selected best views can reasonably present most distinctive regions in the views and contain rich information of the scene, while the worst views exhibit very limited information.
%for us to capture useful information from the worst views.
%
Besides setting the camera to look at the whole scene, we can set it to look at specific areas or distinctive regions in the scene, and find {\em local best views\/}, meaning that we consider only the distinctive regions in the user-specified area when searching for the best view with maximized distinctiveness.
The red and blue boxes in Figure~\ref{fig:large_scale} (a) mark two example areas, while Figures~\ref{fig:large_scale} (e) \& (f) present the corresponding local best views found in the areas.
\xz{Please refer to Supplementary Material Part I for more view selection results.}

%% file: conclusion.tex
\section{Conclusion and Future Work}
\label{sec:conclusion}

\begin{figure}[!t]
	\centering
	\includegraphics[width=0.75\linewidth]{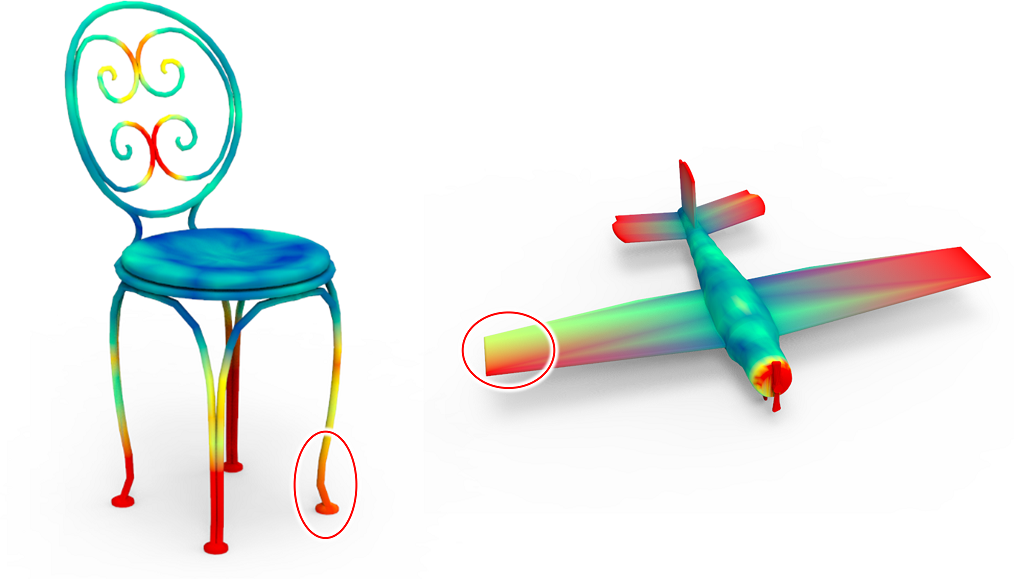}
	\vspace{-3mm}
	\caption{Asymmetry in the detected distinctiveness.}
	\label{fig:limitation}
	\vspace{-2mm}
\end{figure}

We presented a technique to learn and detect distinctive regions on 3D shapes. The technique analyzes a given set of objects {\em without any supervision\/}. The shapes are represented by point clouds, and the analysis is performed by a deep neural network that learns per-point and per-shape features from the point clouds.  Further, we formulate the unsupervised joint loss for a shape clustering task of the per-shape features, thereby implicitly encouraging the network to learn the distinctiveness of the per-point features relative to the shapes in different clusters.
We demonstrated the effectiveness of our method via extensive experiments, and presented several applications based on the network-predicted distinctiveness.

Despite the promising performance that our method has achieved, \NEW{one limitation is that insufficient training samples or severely uneven class size would certainly affect the network's classification capability.
However, such requirement on training data also appears in typical deep-learning-based methods.}
%one major limitation of our method is the asymmetrical distinctive regions detected on shapes; see Figure~\ref{fig:limitation}.
On the other hand, our method may detect asymmetrical distinctive regions on shapes; see Figure~\ref{fig:limitation}.
% on several shapes; see Figure~\ref{fig:limitation}.
This may be related to our current distinctiveness extraction method, which is rather local when extracting the per-point distinctiveness.
In the future, we plan to integrate the prior knowledge of shape symmetry into our current analysis framework when extracting the distinctiveness.
\NEW{While distinction and saliency are different (as explained in the introduction), we may further explore their relationship by taking a machine learning approach.}
Besides, armed with the distinctiveness analysis, in our future work, we are considering generating novel shapes with control on their distinctive features, making them more inter-class distinct and possibly more intra-class distinct, thereby enriching the variability within the class, while remaining distinct to the other classes. Generally speaking, we believe that a stronger and better set analysis will lead, in the future, to better synthesis of 3D shapes.
%\dc{I hope I do not propose here a too attractive idea.}
%\phil{but certainly this is a very good direction, though difficult to realize in a short period of time}

%\phil{TODO: after you finished the supp., please double check all the supp. material index inside the main paper}

%% file: distinctive-points.bbl

\begin{thebibliography}{58}

%%% the Web address.

\ifx \showCODEN    \undefined \def \showCODEN     #1{\unskip}     \fi
\ifx \showDOI      \undefined \def \showDOI       #1{#1}\fi
\ifx \showISBNx    \undefined \def \showISBNx     #1{\unskip}     \fi
\ifx \showISBNxiii \undefined \def \showISBNxiii  #1{\unskip}     \fi
\ifx \showISSN     \undefined \def \showISSN      #1{\unskip}     \fi
\ifx \showLCCN     \undefined \def \showLCCN      #1{\unskip}     \fi
\ifx \shownote     \undefined \def \shownote      #1{#1}          \fi
\ifx \showarticletitle \undefined \def \showarticletitle #1{#1}   \fi
\ifx \showURL      \undefined \def \showURL       {\relax}        \fi
% invisible to TeX
\providecommand\bibfield[2]{#2}
\providecommand\bibinfo[2]{#2}
\providecommand\natexlab[1]{#1}
\providecommand\showeprint[2][]{arXiv:#2}

\bibitem[\protect\citeauthoryear{Abadi, Barham, Chen, Chen, Davis, Dean, Devin,
  Ghemawat, Irving, Isard, Kudlur, Levenberg, Monga, Moore, Murray, Steiner,
  Tucker, Vasudevan, Warden, Wicke, Yu, and Zheng}{Abadi et~al\mbox{.}}{2016}]%
        {45381}
\bibfield{author}{\bibinfo{person}{Martin Abadi}, \bibinfo{person}{Paul
  Barham}, \bibinfo{person}{Jianmin Chen}, \bibinfo{person}{Zhifeng Chen},
  \bibinfo{person}{Andy Davis}, \bibinfo{person}{Jeffrey Dean},
  \bibinfo{person}{Matthieu Devin}, \bibinfo{person}{Sanjay Ghemawat},
  \bibinfo{person}{Geoffrey Irving}, \bibinfo{person}{Michael Isard},
  \bibinfo{person}{Manjunath Kudlur}, \bibinfo{person}{Josh Levenberg},
  \bibinfo{person}{Rajat Monga}, \bibinfo{person}{Sherry Moore},
  \bibinfo{person}{Derek~G. Murray}, \bibinfo{person}{Benoit Steiner},
  \bibinfo{person}{Paul Tucker}, \bibinfo{person}{Vijay Vasudevan},
  \bibinfo{person}{Pete Warden}, \bibinfo{person}{Martin Wicke},
  \bibinfo{person}{Yuan Yu}, {and} \bibinfo{person}{Xiaoqiang Zheng}.}
  \bibinfo{year}{2016}\natexlab{}.
\newblock \showarticletitle{TensorFlow: A system for large-scale machine
  learning}. In \bibinfo{booktitle}{\emph{12th USENIX Symposium on Operating
  Systems Design and Implementation (OSDI 16)}}. \bibinfo{pages}{265--283}.
\newblock
\urldef\tempurl%
\url{https://www.usenix.org/system/files/conference/osdi16/osdi16-abadi.pdf}
\showURL{%
\tempurl}


\bibitem[\protect\citeauthoryear{Alliez, Meyer, and Desbrun}{Alliez
  et~al\mbox{.}}{2002}]%
        {alliez2002interactive}
\bibfield{author}{\bibinfo{person}{Pierre Alliez}, \bibinfo{person}{Mark
  Meyer}, {and} \bibinfo{person}{Mathieu Desbrun}.}
  \bibinfo{year}{2002}\natexlab{}.
\newblock \showarticletitle{\NEW{Interactive geometry remeshing}}.
\newblock \bibinfo{journal}{\emph{ACM Transactions on Graphics (SIGGRAPH)}}
  \bibinfo{volume}{21}, \bibinfo{number}{3} (\bibinfo{year}{2002}),
  \bibinfo{pages}{347--354}.
\newblock


\bibitem[\protect\citeauthoryear{Ancona, Ceolini, Oztireli, and Gross}{Ancona
  et~al\mbox{.}}{2018}]%
        {ancona2018towards}
\bibfield{author}{\bibinfo{person}{Marco Ancona}, \bibinfo{person}{Enea
  Ceolini}, \bibinfo{person}{Cengiz Oztireli}, {and} \bibinfo{person}{Markus
  Gross}.} \bibinfo{year}{2018}\natexlab{}.
\newblock \showarticletitle{Towards better understanding of gradient-based
  attribution methods for deep neural networks}. In
  \bibinfo{booktitle}{\emph{International Conference on Learning
  Representations (ICLR)}}.
\newblock


\bibitem[\protect\citeauthoryear{Aoki, Goforth, Srivatsan, and Lucey}{Aoki
  et~al\mbox{.}}{2019}]%
        {aoki2019pointnetlk}
\bibfield{author}{\bibinfo{person}{Yasuhiro Aoki}, \bibinfo{person}{Hunter
  Goforth}, \bibinfo{person}{Rangaprasad~Arun Srivatsan}, {and}
  \bibinfo{person}{Simon Lucey}.} \bibinfo{year}{2019}\natexlab{}.
\newblock \showarticletitle{\NEW{{PointNetLK}: Robust \& efficient point cloud
  registration using {PointNet}}}. In \bibinfo{booktitle}{\emph{IEEE Conference
  on Computer Vision and Pattern Recognition (CVPR)}}.
  \bibinfo{pages}{7163--7172}.
\newblock


\bibitem[\protect\citeauthoryear{Bachman, Hjelm, and Buchwalter}{Bachman
  et~al\mbox{.}}{2019}]%
        {bachman2019learning}
\bibfield{author}{\bibinfo{person}{Philip Bachman}, \bibinfo{person}{R~Devon
  Hjelm}, {and} \bibinfo{person}{William Buchwalter}.}
  \bibinfo{year}{2019}\natexlab{}.
\newblock \showarticletitle{\NEW{Learning representations by maximizing mutual
  information across views}}. In \bibinfo{booktitle}{\emph{International
  Conference on Neural Information Processing Systems (NIPS)}}.
  \bibinfo{pages}{15509--15519}.
\newblock


\bibitem[\protect\citeauthoryear{Castellani, Cristani, Fantoni, and
  Murino}{Castellani et~al\mbox{.}}{2008}]%
        {castellani2008sparse}
\bibfield{author}{\bibinfo{person}{Umberto Castellani}, \bibinfo{person}{Marco
  Cristani}, \bibinfo{person}{Simone Fantoni}, {and} \bibinfo{person}{Vittorio
  Murino}.} \bibinfo{year}{2008}\natexlab{}.
\newblock \showarticletitle{Sparse points matching by combining 3{D} mesh
  saliency with statistical descriptors}.
\newblock \bibinfo{journal}{\emph{Computer Graphics Forum (Eurographics)}}
  \bibinfo{volume}{27}, \bibinfo{number}{2} (\bibinfo{year}{2008}),
  \bibinfo{pages}{643--652}.
\newblock


\bibitem[\protect\citeauthoryear{Chen, Saparov, Pang, and Funkhouser}{Chen
  et~al\mbox{.}}{2012}]%
        {Chen:2012:SPO}
\bibfield{author}{\bibinfo{person}{Xiaobai Chen}, \bibinfo{person}{Abulhair
  Saparov}, \bibinfo{person}{Bill Pang}, {and} \bibinfo{person}{Thomas
  Funkhouser}.} \bibinfo{year}{2012}\natexlab{}.
\newblock \showarticletitle{Schelling points on 3{D} surface meshes}.
\newblock \bibinfo{journal}{\emph{ACM Transactions on Graphics (SIGGRAPH)}}
  \bibinfo{volume}{31}, \bibinfo{number}{4} (\bibinfo{year}{2012}),
  \bibinfo{pages}{29:1--29:12}.
\newblock


\bibitem[\protect\citeauthoryear{Doersch, Singh, Gupta, Sivic, and
  Efros}{Doersch et~al\mbox{.}}{2012}]%
        {doersch2012makes}
\bibfield{author}{\bibinfo{person}{Carl Doersch}, \bibinfo{person}{Saurabh
  Singh}, \bibinfo{person}{Abhinav Gupta}, \bibinfo{person}{Josef Sivic}, {and}
  \bibinfo{person}{Alexei Efros}.} \bibinfo{year}{2012}\natexlab{}.
\newblock \showarticletitle{What makes {P}aris look like {P}aris?}
\newblock \bibinfo{journal}{\emph{ACM Transactions on Graphics (SIGGRAPH)}}
  \bibinfo{volume}{31}, \bibinfo{number}{4} (\bibinfo{year}{2012}),
  \bibinfo{pages}{101:1--101:9}.
\newblock


\bibitem[\protect\citeauthoryear{Dutagaci, Cheung, and Godil}{Dutagaci
  et~al\mbox{.}}{2012}]%
        {dutagaci2012evaluation}
\bibfield{author}{\bibinfo{person}{Helin Dutagaci}, \bibinfo{person}{Chun~Pan
  Cheung}, {and} \bibinfo{person}{Afzal Godil}.}
  \bibinfo{year}{2012}\natexlab{}.
\newblock \showarticletitle{Evaluation of 3D interest point detection
  techniques via human-generated ground truth}.
\newblock \bibinfo{journal}{\emph{The Visual Computer}} \bibinfo{volume}{28},
  \bibinfo{number}{9} (\bibinfo{year}{2012}), \bibinfo{pages}{901--917}.
\newblock


\bibitem[\protect\citeauthoryear{Gal and Cohen-Or}{Gal and Cohen-Or}{2006}]%
        {gal2006salient}
\bibfield{author}{\bibinfo{person}{Ran Gal} {and} \bibinfo{person}{Daniel
  Cohen-Or}.} \bibinfo{year}{2006}\natexlab{}.
\newblock \showarticletitle{Salient geometric features for partial shape
  matching and similarity}.
\newblock \bibinfo{journal}{\emph{ACM Transactions on Graphics}}
  \bibinfo{volume}{25}, \bibinfo{number}{1} (\bibinfo{year}{2006}),
  \bibinfo{pages}{130--150}.
\newblock


\bibitem[\protect\citeauthoryear{Garland and Heckbert}{Garland and
  Heckbert}{1997}]%
        {garland1997surface}
\bibfield{author}{\bibinfo{person}{Michael Garland} {and}
  \bibinfo{person}{Paul~S. Heckbert}.} \bibinfo{year}{1997}\natexlab{}.
\newblock \showarticletitle{\NEW{Surface simplification using quadric error
  metrics}}. In \bibinfo{booktitle}{\emph{Proceedings of SIGGRAPH}}.
  \bibinfo{pages}{209--216}.
\newblock


\bibitem[\protect\citeauthoryear{Hadsell, Chopra, and LeCun}{Hadsell
  et~al\mbox{.}}{2006}]%
        {hadsell2006dimensionality}
\bibfield{author}{\bibinfo{person}{Raia Hadsell}, \bibinfo{person}{Sumit
  Chopra}, {and} \bibinfo{person}{Yann LeCun}.}
  \bibinfo{year}{2006}\natexlab{}.
\newblock \showarticletitle{Dimensionality reduction by learning an invariant
  mapping}. In \bibinfo{booktitle}{\emph{IEEE Conference on Computer Vision and
  Pattern Recognition (CVPR)}}. \bibinfo{pages}{1735--1742}.
\newblock


\bibitem[\protect\citeauthoryear{H{\'e}naff, Razavi, Doersch, Eslami, and
  Oord}{H{\'e}naff et~al\mbox{.}}{2019}]%
        {henaff2019data}
\bibfield{author}{\bibinfo{person}{Olivier~J. H{\'e}naff}, \bibinfo{person}{Ali
  Razavi}, \bibinfo{person}{Carl Doersch}, \bibinfo{person}{S.~M. Eslami},
  {and} \bibinfo{person}{Aaron van~den Oord}.} \bibinfo{year}{2019}\natexlab{}.
\newblock \showarticletitle{\NEW{Data-efficient image recognition with
  contrastive predictive coding}}.
\newblock \bibinfo{journal}{\emph{arXiv preprint arXiv:1905.09272}}
  (\bibinfo{year}{2019}).
\newblock


\bibitem[\protect\citeauthoryear{Hinton, Vinyals, and Dean}{Hinton
  et~al\mbox{.}}{2015}]%
        {hinton2015distilling}
\bibfield{author}{\bibinfo{person}{Geoffrey Hinton}, \bibinfo{person}{Oriol
  Vinyals}, {and} \bibinfo{person}{Jeff Dean}.}
  \bibinfo{year}{2015}\natexlab{}.
\newblock \showarticletitle{Distilling the knowledge in a neural network}.
\newblock \bibinfo{journal}{\emph{arXiv preprint arXiv:1503.02531}}
  (\bibinfo{year}{2015}).
\newblock


\bibitem[\protect\citeauthoryear{Hjelm, Fedorov, Lavoie-Marchildon, Grewal,
  Bachman, Trischler, and Bengio}{Hjelm et~al\mbox{.}}{2019}]%
        {hjelm2018learning}
\bibfield{author}{\bibinfo{person}{R~Devon Hjelm}, \bibinfo{person}{Alex
  Fedorov}, \bibinfo{person}{Samuel Lavoie-Marchildon}, \bibinfo{person}{Karan
  Grewal}, \bibinfo{person}{Phil Bachman}, \bibinfo{person}{Adam Trischler},
  {and} \bibinfo{person}{Yoshua Bengio}.} \bibinfo{year}{2019}\natexlab{}.
\newblock \showarticletitle{\NEW{Learning deep representations by mutual
  information estimation and maximization}}. In
  \bibinfo{booktitle}{\emph{International Conference on Learning
  Representations (ICLR)}}.
\newblock


\bibitem[\protect\citeauthoryear{Hua, Tran, and Yeung}{Hua
  et~al\mbox{.}}{2018}]%
        {hua2018pointwise}
\bibfield{author}{\bibinfo{person}{Binh-Son Hua}, \bibinfo{person}{Minh-Khoi
  Tran}, {and} \bibinfo{person}{Sai-Kit Yeung}.}
  \bibinfo{year}{2018}\natexlab{}.
\newblock \showarticletitle{Pointwise convolutional neural networks}. In
  \bibinfo{booktitle}{\emph{IEEE Conference on Computer Vision and Pattern
  Recognition (CVPR)}}. \bibinfo{pages}{984--993}.
\newblock


\bibitem[\protect\citeauthoryear{Juneja, Vedaldi, Jawahar, and
  Zisserman}{Juneja et~al\mbox{.}}{2013}]%
        {juneja2013blocks}
\bibfield{author}{\bibinfo{person}{Mayank Juneja}, \bibinfo{person}{Andrea
  Vedaldi}, \bibinfo{person}{C.V. Jawahar}, {and} \bibinfo{person}{Andrew
  Zisserman}.} \bibinfo{year}{2013}\natexlab{}.
\newblock \showarticletitle{Blocks that shout: Distinctive parts for scene
  classification}. In \bibinfo{booktitle}{\emph{IEEE Conference on Computer
  Vision and Pattern Recognition (CVPR)}}. \bibinfo{pages}{923--930}.
\newblock


\bibitem[\protect\citeauthoryear{Kingma and Ba}{Kingma and Ba}{2014}]%
        {kingma2014adam}
\bibfield{author}{\bibinfo{person}{Diederik~P. Kingma} {and}
  \bibinfo{person}{Jimmy Ba}.} \bibinfo{year}{2014}\natexlab{}.
\newblock \showarticletitle{Adam: A method for stochastic optimization}.
\newblock \bibinfo{journal}{\emph{arXiv preprint arXiv:1412.6980}}
  (\bibinfo{year}{2014}).
\newblock


\bibitem[\protect\citeauthoryear{Le and Duan}{Le and Duan}{2018}]%
        {le2018pointgrid}
\bibfield{author}{\bibinfo{person}{Truc Le} {and} \bibinfo{person}{Ye Duan}.}
  \bibinfo{year}{2018}\natexlab{}.
\newblock \showarticletitle{PointGrid: A deep network for 3D shape
  understanding}. In \bibinfo{booktitle}{\emph{IEEE Conference on Computer
  Vision and Pattern Recognition (CVPR)}}. \bibinfo{pages}{9204--9214}.
\newblock


\bibitem[\protect\citeauthoryear{Lee, Varshney, and Jacobs}{Lee
  et~al\mbox{.}}{2005}]%
        {lee2005mesh}
\bibfield{author}{\bibinfo{person}{Chang~Ha Lee}, \bibinfo{person}{Amitabh
  Varshney}, {and} \bibinfo{person}{David~W. Jacobs}.}
  \bibinfo{year}{2005}\natexlab{}.
\newblock \showarticletitle{Mesh saliency}.
\newblock \bibinfo{journal}{\emph{ACM Transactions on Graphics (SIGGRAPH)}}
  \bibinfo{volume}{24}, \bibinfo{number}{3} (\bibinfo{year}{2005}),
  \bibinfo{pages}{659--666}.
\newblock


\bibitem[\protect\citeauthoryear{Leifman, Shtrom, and Tal}{Leifman
  et~al\mbox{.}}{2012}]%
        {leifman2012surface}
\bibfield{author}{\bibinfo{person}{George Leifman}, \bibinfo{person}{Elizabeth
  Shtrom}, {and} \bibinfo{person}{Ayellet Tal}.}
  \bibinfo{year}{2012}\natexlab{}.
\newblock \showarticletitle{Surface regions of interest for viewpoint
  selection}. In \bibinfo{booktitle}{\emph{IEEE Conference on Computer Vision
  and Pattern Recognition (CVPR)}}. \bibinfo{pages}{414--421}.
\newblock


\bibitem[\protect\citeauthoryear{Li, Bu, Sun, Wu, Di, and Chen}{Li
  et~al\mbox{.}}{2018}]%
        {li2018pointcnn}
\bibfield{author}{\bibinfo{person}{Yangyan Li}, \bibinfo{person}{Rui Bu},
  \bibinfo{person}{Mingchao Sun}, \bibinfo{person}{Wei Wu},
  \bibinfo{person}{Xinhan Di}, {and} \bibinfo{person}{Baoquan Chen}.}
  \bibinfo{year}{2018}\natexlab{}.
\newblock \showarticletitle{{PointCNN}: Convolution on
  $\mathcal{X}$-transformed points}. In \bibinfo{booktitle}{\emph{International
  Conference on Neural Information Processing Systems (NIPS)}}.
  \bibinfo{pages}{828--838}.
\newblock


\bibitem[\protect\citeauthoryear{Liu, Song, Shao, Jin, and Wang}{Liu
  et~al\mbox{.}}{2018}]%
        {liu2018transductive}
\bibfield{author}{\bibinfo{person}{Yu Liu}, \bibinfo{person}{Guanglu Song},
  \bibinfo{person}{Jing Shao}, \bibinfo{person}{Xiao Jin}, {and}
  \bibinfo{person}{Xiaogang Wang}.} \bibinfo{year}{2018}\natexlab{}.
\newblock \showarticletitle{Transductive centroid projection for
  semi-supervised large-scale recognition}. In
  \bibinfo{booktitle}{\emph{European Conference on Computer Vision (ECCV)}}.
  \bibinfo{pages}{70--86}.
\newblock


\bibitem[\protect\citeauthoryear{Ponjou~Tasse, Kosinka, and
  Dodgson}{Ponjou~Tasse et~al\mbox{.}}{2015}]%
        {ponjou2015cluster}
\bibfield{author}{\bibinfo{person}{Flora Ponjou~Tasse}, \bibinfo{person}{Jiri
  Kosinka}, {and} \bibinfo{person}{Neil Dodgson}.}
  \bibinfo{year}{2015}\natexlab{}.
\newblock \showarticletitle{Cluster-based point set saliency}. In
  \bibinfo{booktitle}{\emph{IEEE International Conference on Computer Vision
  (ICCV)}}. \bibinfo{pages}{163--171}.
\newblock


\bibitem[\protect\citeauthoryear{Qi, Su, Mo, and Guibas}{Qi
  et~al\mbox{.}}{2017a}]%
        {qi2016pointnet}
\bibfield{author}{\bibinfo{person}{Charles~R. Qi}, \bibinfo{person}{Hao Su},
  \bibinfo{person}{Kaichun Mo}, {and} \bibinfo{person}{Leonidas~J. Guibas}.}
  \bibinfo{year}{2017}\natexlab{a}.
\newblock \showarticletitle{{PointNet}: Deep learning on point sets for {3D}
  classification and segmentation}. In \bibinfo{booktitle}{\emph{IEEE
  Conference on Computer Vision and Pattern Recognition (CVPR)}}.
  \bibinfo{pages}{652--660}.
\newblock


\bibitem[\protect\citeauthoryear{Qi, Yi, Su, and Guibas}{Qi
  et~al\mbox{.}}{2017b}]%
        {qi2017pointnet++}
\bibfield{author}{\bibinfo{person}{Charles~R. Qi}, \bibinfo{person}{Li Yi},
  \bibinfo{person}{Hao Su}, {and} \bibinfo{person}{Leonidas~J. Guibas}.}
  \bibinfo{year}{2017}\natexlab{b}.
\newblock \showarticletitle{Point{N}et++: Deep hierarchical feature learning on
  point sets in a metric space}. In \bibinfo{booktitle}{\emph{International
  Conference on Neural Information Processing Systems (NIPS)}}.
  \bibinfo{pages}{5099--5108}.
\newblock


\bibitem[\protect\citeauthoryear{Selvaraju, Cogswell, Das, Vedantam, Parikh,
  and Batra}{Selvaraju et~al\mbox{.}}{2017}]%
        {Selvaraju2017grad}
\bibfield{author}{\bibinfo{person}{Ramprasaath~R. Selvaraju},
  \bibinfo{person}{Michael Cogswell}, \bibinfo{person}{Abhishek Das},
  \bibinfo{person}{Ramakrishna Vedantam}, \bibinfo{person}{Devi Parikh}, {and}
  \bibinfo{person}{Dhruv Batra}.} \bibinfo{year}{2017}\natexlab{}.
\newblock \showarticletitle{Grad-CAM: Visual explanations from deep networks
  via gradient-based localization}. In \bibinfo{booktitle}{\emph{IEEE
  International Conference on Computer Vision (ICCV)}}.
  \bibinfo{pages}{618--626}.
\newblock


\bibitem[\protect\citeauthoryear{Shen, Feng, Yang, and Tian}{Shen
  et~al\mbox{.}}{2018}]%
        {shen2018mining}
\bibfield{author}{\bibinfo{person}{Yiru Shen}, \bibinfo{person}{Chen Feng},
  \bibinfo{person}{Yaoqing Yang}, {and} \bibinfo{person}{Dong Tian}.}
  \bibinfo{year}{2018}\natexlab{}.
\newblock \showarticletitle{Mining point cloud local structures by kernel
  correlation and graph pooling}. In \bibinfo{booktitle}{\emph{IEEE Conference
  on Computer Vision and Pattern Recognition (CVPR)}}.
  \bibinfo{pages}{4548--4557}.
\newblock


\bibitem[\protect\citeauthoryear{Shilane and Funkhouser}{Shilane and
  Funkhouser}{2006}]%
        {shilane2006selecting}
\bibfield{author}{\bibinfo{person}{Philip Shilane} {and}
  \bibinfo{person}{Thomas Funkhouser}.} \bibinfo{year}{2006}\natexlab{}.
\newblock \showarticletitle{Selecting distinctive 3D shape descriptors for
  similarity retrieval}. In \bibinfo{booktitle}{\emph{IEEE Intl. Conf. on Shape
  Modeling and Applications (SMI)}}. \bibinfo{pages}{18:1--18:10}.
\newblock


\bibitem[\protect\citeauthoryear{Shilane and Funkhouser}{Shilane and
  Funkhouser}{2007}]%
        {shilane2007distinctive}
\bibfield{author}{\bibinfo{person}{Philip Shilane} {and}
  \bibinfo{person}{Thomas Funkhouser}.} \bibinfo{year}{2007}\natexlab{}.
\newblock \showarticletitle{Distinctive regions of 3{D} surfaces}.
\newblock \bibinfo{journal}{\emph{ACM Transactions on Graphics}}
  \bibinfo{volume}{26}, \bibinfo{number}{2} (\bibinfo{year}{2007}),
  \bibinfo{pages}{7:1--7:15}.
\newblock


\bibitem[\protect\citeauthoryear{Shilane, Min, Kazhdan, and Funkhouser}{Shilane
  et~al\mbox{.}}{2004}]%
        {shilane2004princeton}
\bibfield{author}{\bibinfo{person}{Philip Shilane}, \bibinfo{person}{Patrick
  Min}, \bibinfo{person}{Michael Kazhdan}, {and} \bibinfo{person}{Thomas
  Funkhouser}.} \bibinfo{year}{2004}\natexlab{}.
\newblock \showarticletitle{The {P}rinceton shape benchmark}. In
  \bibinfo{booktitle}{\emph{IEEE Intl. Conf. on Shape Modeling and Applications
  (SMI)}}. \bibinfo{pages}{167--178}.
\newblock


\bibitem[\protect\citeauthoryear{Shrikumar, Greenside, and Kundaje}{Shrikumar
  et~al\mbox{.}}{2017}]%
        {shrikumar2017learning}
\bibfield{author}{\bibinfo{person}{Avanti Shrikumar}, \bibinfo{person}{Peyton
  Greenside}, {and} \bibinfo{person}{Anshul Kundaje}.}
  \bibinfo{year}{2017}\natexlab{}.
\newblock \showarticletitle{Learning Important Features Through Propagating
  Activation Differences}. In \bibinfo{booktitle}{\emph{Proceedings of
  International Conference on Machine Learning (ICML)}}.
  \bibinfo{pages}{3145--3153}.
\newblock


\bibitem[\protect\citeauthoryear{Shtrom, Leifman, and Tal}{Shtrom
  et~al\mbox{.}}{2013}]%
        {shtrom2013saliency}
\bibfield{author}{\bibinfo{person}{Elizabeth Shtrom}, \bibinfo{person}{George
  Leifman}, {and} \bibinfo{person}{Ayellet Tal}.}
  \bibinfo{year}{2013}\natexlab{}.
\newblock \showarticletitle{Saliency detection in large point sets}. In
  \bibinfo{booktitle}{\emph{IEEE International Conference on Computer Vision
  (ICCV)}}. \bibinfo{pages}{3591--3598}.
\newblock


\bibitem[\protect\citeauthoryear{Shu, Xin, Xu, Liu, and Kavan}{Shu
  et~al\mbox{.}}{2019}]%
        {shu2018detecting}
\bibfield{author}{\bibinfo{person}{Zhenyu Shu}, \bibinfo{person}{Shiqing Xin},
  \bibinfo{person}{Xin Xu}, \bibinfo{person}{Ligang Liu}, {and}
  \bibinfo{person}{Ladislav Kavan}.} \bibinfo{year}{2019}\natexlab{}.
\newblock \showarticletitle{Detecting 3{D} points of interest using multiple
  features and stacked auto-encoder}.
\newblock \bibinfo{journal}{\emph{IEEE Transactions Visualization $\&$ Computer
  Graphics}} \bibinfo{volume}{25}, \bibinfo{number}{8} (\bibinfo{year}{2019}),
  \bibinfo{pages}{2583--2596}.
\newblock


\bibitem[\protect\citeauthoryear{Singh, Gupta, and Efros}{Singh
  et~al\mbox{.}}{2012}]%
        {singh2012unsupervised}
\bibfield{author}{\bibinfo{person}{Saurabh Singh}, \bibinfo{person}{Abhinav
  Gupta}, {and} \bibinfo{person}{Alexei~A. Efros}.}
  \bibinfo{year}{2012}\natexlab{}.
\newblock \showarticletitle{Unsupervised discovery of mid-level discriminative
  patches}.
\newblock In \bibinfo{booktitle}{\emph{European Conference on Computer Vision
  (ECCV)}}. \bibinfo{pages}{73--86}.
\newblock


\bibitem[\protect\citeauthoryear{Song, Liu, and Rosin}{Song
  et~al\mbox{.}}{2018}]%
        {song2018distinction}
\bibfield{author}{\bibinfo{person}{Ran Song}, \bibinfo{person}{Yonghuai Liu},
  {and} \bibinfo{person}{Paul Rosin}.} \bibinfo{year}{2018}\natexlab{}.
\newblock \showarticletitle{Distinction of 3D objects and scenes via
  classification network and {M}arkov random field}.
\newblock \bibinfo{journal}{\emph{IEEE Transactions Visualization $\&$ Computer
  Graphics}} (\bibinfo{year}{2018}), \bibinfo{pages}{To appear}.
\newblock


\bibitem[\protect\citeauthoryear{Stella and Shi}{Stella and Shi}{2003}]%
        {stella2003multiclass}
\bibfield{author}{\bibinfo{person}{X.~Yu Stella} {and} \bibinfo{person}{Jianbo
  Shi}.} \bibinfo{year}{2003}\natexlab{}.
\newblock \showarticletitle{Multiclass spectral clustering}. In
  \bibinfo{booktitle}{\emph{IEEE International Conference on Computer Vision
  (ICCV)}}. \bibinfo{pages}{313--320}.
\newblock


\bibitem[\protect\citeauthoryear{Su, Jampani, Sun, Maji, Kalogerakis, Yang, and
  Kautz}{Su et~al\mbox{.}}{2018}]%
        {su2018splatnet}
\bibfield{author}{\bibinfo{person}{Hang Su}, \bibinfo{person}{Varun Jampani},
  \bibinfo{person}{Deqing Sun}, \bibinfo{person}{Subhransu Maji},
  \bibinfo{person}{Evangelos Kalogerakis}, \bibinfo{person}{Ming-Hsuan Yang},
  {and} \bibinfo{person}{Jan Kautz}.} \bibinfo{year}{2018}\natexlab{}.
\newblock \showarticletitle{{SPLATNet}: Sparse lattice networks for point cloud
  processing}. In \bibinfo{booktitle}{\emph{IEEE Conference on Computer Vision
  and Pattern Recognition (CVPR)}}. \bibinfo{pages}{2530--2539}.
\newblock


\bibitem[\protect\citeauthoryear{Sun and Ponce}{Sun and Ponce}{2013}]%
        {sun2013learning}
\bibfield{author}{\bibinfo{person}{Jian Sun} {and} \bibinfo{person}{Jean
  Ponce}.} \bibinfo{year}{2013}\natexlab{}.
\newblock \showarticletitle{Learning discriminative part detectors for image
  classification and cosegmentation}. In \bibinfo{booktitle}{\emph{IEEE
  International Conference on Computer Vision (ICCV)}}.
  \bibinfo{pages}{3400--3407}.
\newblock


\bibitem[\protect\citeauthoryear{Sundararajan, Taly, and Yan}{Sundararajan
  et~al\mbox{.}}{2017}]%
        {sundararajan2017axiomatic}
\bibfield{author}{\bibinfo{person}{Mukund Sundararajan}, \bibinfo{person}{Ankur
  Taly}, {and} \bibinfo{person}{Qiqi Yan}.} \bibinfo{year}{2017}\natexlab{}.
\newblock \showarticletitle{Axiomatic Attribution for Deep Networks}. In
  \bibinfo{booktitle}{\emph{Proceedings of International Conference on Machine
  Learning (ICML)}}. \bibinfo{pages}{3319--3328}.
\newblock


\bibitem[\protect\citeauthoryear{Tangelder and Veltkamp}{Tangelder and
  Veltkamp}{2004}]%
        {tangelder2004survey}
\bibfield{author}{\bibinfo{person}{Johan W.~H. Tangelder} {and}
  \bibinfo{person}{Remco~C. Veltkamp}.} \bibinfo{year}{2004}\natexlab{}.
\newblock \showarticletitle{A survey of content based 3{D} shape retrieval
  methods}. In \bibinfo{booktitle}{\emph{IEEE Intl. Conf. on Shape Modeling and
  Applications (SMI)}}. \bibinfo{pages}{145--156}.
\newblock


\bibitem[\protect\citeauthoryear{Von~Luxburg}{Von~Luxburg}{2007}]%
        {von2007tutorial}
\bibfield{author}{\bibinfo{person}{Ulrike Von~Luxburg}.}
  \bibinfo{year}{2007}\natexlab{}.
\newblock \showarticletitle{A tutorial on spectral clustering}.
\newblock \bibinfo{journal}{\emph{Statistics and computing}}
  \bibinfo{volume}{17}, \bibinfo{number}{4} (\bibinfo{year}{2007}),
  \bibinfo{pages}{395--416}.
\newblock


\bibitem[\protect\citeauthoryear{Wang, Koch, Holmqvist, and Alexa}{Wang
  et~al\mbox{.}}{2018}]%
        {wang2018tracking}
\bibfield{author}{\bibinfo{person}{Xi Wang}, \bibinfo{person}{Sebastian Koch},
  \bibinfo{person}{Kenneth Holmqvist}, {and} \bibinfo{person}{Marc Alexa}.}
  \bibinfo{year}{2018}\natexlab{}.
\newblock \showarticletitle{Tracking the gaze on objects in 3{D}: how do people
  really look at the {B}unny?}
\newblock \bibinfo{journal}{\emph{ACM Transactions on Graphics (SIGGRAPH
  Asia)}} \bibinfo{volume}{37}, \bibinfo{number}{6} (\bibinfo{year}{2018}),
  \bibinfo{pages}{188:1--188:18}.
\newblock


\bibitem[\protect\citeauthoryear{Wang, Choi, Morariu, and Davis}{Wang
  et~al\mbox{.}}{2016}]%
        {wang2016mining}
\bibfield{author}{\bibinfo{person}{Yaming Wang}, \bibinfo{person}{Jonghyun
  Choi}, \bibinfo{person}{Vlad Morariu}, {and} \bibinfo{person}{Larry~S.
  Davis}.} \bibinfo{year}{2016}\natexlab{}.
\newblock \showarticletitle{Mining discriminative triplets of patches for
  fine-grained classification}. In \bibinfo{booktitle}{\emph{IEEE Conference on
  Computer Vision and Pattern Recognition (CVPR)}}.
  \bibinfo{pages}{1163--1172}.
\newblock


\bibitem[\protect\citeauthoryear{Wang and Solomon}{Wang and Solomon}{2019}]%
        {wang2019deep}
\bibfield{author}{\bibinfo{person}{Yue Wang} {and} \bibinfo{person}{Justin~M.
  Solomon}.} \bibinfo{year}{2019}\natexlab{}.
\newblock \showarticletitle{\NEW{Deep Closest Point: Learning Representations
  for Point Cloud Registration}}. In \bibinfo{booktitle}{\emph{IEEE
  International Conference on Computer Vision (ICCV)}}.
  \bibinfo{pages}{3523--3532}.
\newblock


\bibitem[\protect\citeauthoryear{Wang, Sun, Liu, Sarma, Bronstein, and
  Solomon}{Wang et~al\mbox{.}}{2019}]%
        {wang2018dynamic}
\bibfield{author}{\bibinfo{person}{Yue Wang}, \bibinfo{person}{Yongbin Sun},
  \bibinfo{person}{Ziwei Liu}, \bibinfo{person}{Sanjay~E. Sarma},
  \bibinfo{person}{Michael~M. Bronstein}, {and} \bibinfo{person}{Justin~M.
  Solomon}.} \bibinfo{year}{2019}\natexlab{}.
\newblock \showarticletitle{Dynamic graph CNN for learning on point clouds}.
\newblock \bibinfo{journal}{\emph{ACM Transactions on Graphics}}
  \bibinfo{volume}{38}, \bibinfo{number}{5} (\bibinfo{year}{2019}),
  \bibinfo{pages}{146:1--146:12}.
\newblock


\bibitem[\protect\citeauthoryear{Warehouse}{Warehouse}{2019}]%
        {warehouse}
\bibfield{author}{\bibinfo{person}{3D Warehouse}.}
  \bibinfo{year}{2019}\natexlab{}.
\newblock
\newblock
\urldef\tempurl%
\url{https://3dwarehouse.sketchup.com/}
\showURL{%
\tempurl}
\newblock
\shownote{[Online; accessed 02-Jan-2019].}


\bibitem[\protect\citeauthoryear{Wen, Zhang, Li, and Qiao}{Wen
  et~al\mbox{.}}{2016}]%
        {wen2016discriminative}
\bibfield{author}{\bibinfo{person}{Yandong Wen}, \bibinfo{person}{Kaipeng
  Zhang}, \bibinfo{person}{Zhifeng Li}, {and} \bibinfo{person}{Yu Qiao}.}
  \bibinfo{year}{2016}\natexlab{}.
\newblock \showarticletitle{A discriminative feature learning approach for deep
  face recognition}. In \bibinfo{booktitle}{\emph{European Conference on
  Computer Vision (ECCV)}}. \bibinfo{pages}{499--515}.
\newblock


\bibitem[\protect\citeauthoryear{Woo, Park, Lee, and So~Kweon}{Woo
  et~al\mbox{.}}{2018}]%
        {woo2018cbam}
\bibfield{author}{\bibinfo{person}{Sanghyun Woo}, \bibinfo{person}{Jongchan
  Park}, \bibinfo{person}{Joon-Young Lee}, {and} \bibinfo{person}{In
  So~Kweon}.} \bibinfo{year}{2018}\natexlab{}.
\newblock \showarticletitle{{CBAM}: Convolutional block attention module}. In
  \bibinfo{booktitle}{\emph{European Conference on Computer Vision (ECCV)}}.
  \bibinfo{pages}{3--19}.
\newblock


\bibitem[\protect\citeauthoryear{Wu, Song, Khosla, Yu, Zhang, Tang, and
  Xiao}{Wu et~al\mbox{.}}{2015}]%
        {wu20153d}
\bibfield{author}{\bibinfo{person}{Zhirong Wu}, \bibinfo{person}{Shuran Song},
  \bibinfo{person}{Aditya Khosla}, \bibinfo{person}{Fisher Yu},
  \bibinfo{person}{Linguang Zhang}, \bibinfo{person}{Xiaoou Tang}, {and}
  \bibinfo{person}{Jianxiong Xiao}.} \bibinfo{year}{2015}\natexlab{}.
\newblock \showarticletitle{{3D ShapeNets}: A deep representation for
  volumetric shapes}. In \bibinfo{booktitle}{\emph{IEEE Conference on Computer
  Vision and Pattern Recognition (CVPR)}}. \bibinfo{pages}{1912--1920}.
\newblock


\bibitem[\protect\citeauthoryear{Wu, Xiong, Stella, and Lin}{Wu
  et~al\mbox{.}}{2018}]%
        {wu2018unsupervised}
\bibfield{author}{\bibinfo{person}{Zhirong Wu}, \bibinfo{person}{Yuanjun
  Xiong}, \bibinfo{person}{X.~Yu Stella}, {and} \bibinfo{person}{Dahua Lin}.}
  \bibinfo{year}{2018}\natexlab{}.
\newblock \showarticletitle{Unsupervised feature learning via non-parametric
  instance discrimination}. In \bibinfo{booktitle}{\emph{IEEE Conference on
  Computer Vision and Pattern Recognition (CVPR)}}.
  \bibinfo{pages}{3733--3742}.
\newblock


\bibitem[\protect\citeauthoryear{Xu, Fan, Xu, Zeng, and Qiao}{Xu
  et~al\mbox{.}}{2018}]%
        {xu2018spidercnn}
\bibfield{author}{\bibinfo{person}{Yifan Xu}, \bibinfo{person}{Tianqi Fan},
  \bibinfo{person}{Mingye Xu}, \bibinfo{person}{Long Zeng}, {and}
  \bibinfo{person}{Yu Qiao}.} \bibinfo{year}{2018}\natexlab{}.
\newblock \showarticletitle{{SpiderCNN}: Deep learning on point sets with
  parameterized convolutional filters}. In \bibinfo{booktitle}{\emph{European
  Conference on Computer Vision (ECCV)}}. \bibinfo{pages}{90--105}.
\newblock


\bibitem[\protect\citeauthoryear{Yang, Feng, Shen, and Tian}{Yang
  et~al\mbox{.}}{2018}]%
        {yang2018foldingnet}
\bibfield{author}{\bibinfo{person}{Yaoqing Yang}, \bibinfo{person}{Chen Feng},
  \bibinfo{person}{Yiru Shen}, {and} \bibinfo{person}{Dong Tian}.}
  \bibinfo{year}{2018}\natexlab{}.
\newblock \showarticletitle{{FoldingNet}: Point cloud auto-encoder via deep
  grid deformation}. In \bibinfo{booktitle}{\emph{IEEE Conference on Computer
  Vision and Pattern Recognition (CVPR)}}. \bibinfo{pages}{206--215}.
\newblock


\bibitem[\protect\citeauthoryear{Yu, Li, Fu, Cohen-Or, and Heng}{Yu
  et~al\mbox{.}}{2018}]%
        {yu-2018-EC-net}
\bibfield{author}{\bibinfo{person}{Lequan Yu}, \bibinfo{person}{Xianzhi Li},
  \bibinfo{person}{Chi-Wing Fu}, \bibinfo{person}{Daniel Cohen-Or}, {and}
  \bibinfo{person}{Pheng-Ann Heng}.} \bibinfo{year}{2018}\natexlab{}.
\newblock \showarticletitle{EC-Net: an Edge-aware Point set Consolidation
  Network}. In \bibinfo{booktitle}{\emph{European Conference on Computer Vision
  (ECCV)}}. \bibinfo{pages}{398--414}.
\newblock


\bibitem[\protect\citeauthoryear{Zeiler and Fergus}{Zeiler and Fergus}{2014}]%
        {zeiler2014visualizing}
\bibfield{author}{\bibinfo{person}{Matthew~D. Zeiler} {and}
  \bibinfo{person}{Rob Fergus}.} \bibinfo{year}{2014}\natexlab{}.
\newblock \showarticletitle{Visualizing and understanding convolutional
  networks}. In \bibinfo{booktitle}{\emph{European Conference on Computer
  Vision (ECCV)}}. \bibinfo{pages}{818--833}.
\newblock


\bibitem[\protect\citeauthoryear{Zhang, Bargal, Lin, Brandt, Shen, and
  Sclaroff}{Zhang et~al\mbox{.}}{2018}]%
        {zhang2018top}
\bibfield{author}{\bibinfo{person}{Jianming Zhang}, \bibinfo{person}{Sarah~Adel
  Bargal}, \bibinfo{person}{Zhe Lin}, \bibinfo{person}{Jonathan Brandt},
  \bibinfo{person}{Xiaohui Shen}, {and} \bibinfo{person}{Stan Sclaroff}.}
  \bibinfo{year}{2018}\natexlab{}.
\newblock \showarticletitle{Top-down neural attention by excitation backprop}.
\newblock \bibinfo{journal}{\emph{International Journal Computer Vision}}
  \bibinfo{volume}{126}, \bibinfo{number}{10} (\bibinfo{year}{2018}),
  \bibinfo{pages}{1084--1102}.
\newblock


\bibitem[\protect\citeauthoryear{Zhang, Le, Panotopoulou, Whiting, and
  Wang}{Zhang et~al\mbox{.}}{2015}]%
        {zhang2015perceptual}
\bibfield{author}{\bibinfo{person}{Xiaoting Zhang}, \bibinfo{person}{Xinyi Le},
  \bibinfo{person}{Athina Panotopoulou}, \bibinfo{person}{Emily Whiting}, {and}
  \bibinfo{person}{Charlie C.~L. Wang}.} \bibinfo{year}{2015}\natexlab{}.
\newblock \showarticletitle{\NEW{Perceptual models of preference in 3{D}
  printing direction}}.
\newblock \bibinfo{journal}{\emph{ACM Transactions on Graphics (SIGGRAPH
  Asia)}} \bibinfo{volume}{34}, \bibinfo{number}{6} (\bibinfo{year}{2015}),
  \bibinfo{pages}{215:1--215:12}.
\newblock


\bibitem[\protect\citeauthoryear{Zhou, Khosla, Lapedriza, Oliva, and
  Torralba}{Zhou et~al\mbox{.}}{2016}]%
        {zhou2016learning}
\bibfield{author}{\bibinfo{person}{Bolei Zhou}, \bibinfo{person}{Aditya
  Khosla}, \bibinfo{person}{Agata Lapedriza}, \bibinfo{person}{Aude Oliva},
  {and} \bibinfo{person}{Antonio Torralba}.} \bibinfo{year}{2016}\natexlab{}.
\newblock \showarticletitle{Learning deep features for discriminative
  localization}. In \bibinfo{booktitle}{\emph{IEEE Conference on Computer
  Vision and Pattern Recognition (CVPR)}}. \bibinfo{pages}{2921--2929}.
\newblock


\end{thebibliography}
